\documentclass[10pt,twocolumn,twoside]{IEEEtran}

\usepackage{graphicx}  

\usepackage{psfrag}    

\usepackage{subfigure} 
\begin{document}
%
\title{Adaptive Langevin Sampler for Separation of \textit{t}-distribution Modelled Astrophysical Maps}
%
%
\author{Koray~Kayabol,~\IEEEmembership{Member,~IEEE,}
        Ercan~E.~Kuruo\u{g}lu,~\IEEEmembership{Senior Member, ~IEEE,}
        Jos\'e~Luis~Sanz,
        B\"{u}lent~Sankur,~\IEEEmembership{Senior Member, ~IEEE,}
        Emanuele~Salerno and~Diego~Herranz
\thanks{Manuscript received September 29, 2009; revised March 15, 2010 and accepted March 28, 2010.
        Koray Kayabol undertook this work with
the support of the "ICTP Programme for Training and Research in
Italian Laboratories, Trieste, Italy, through a specific
operational agreement with CNR-ISTI, Italy. Partial support has
also been given by the Italian Space Agency (ASI), under project
COFIS (Cosmology and Fundamental Physics). The project is
partially supported by CNR-CSIC bilateral project.}
\thanks{K. Kayabol, E. E. Kuruo\u{g}lu and E. Salerno are with the ISTI, CNR, via G. Moruzzi 1,
 56124, Pisa, Italy,
(e-mail: koray.kayabol@isti.cnr.it; ercan.kuruoglu@isti.cnr.it;
emanuele.salerno@isti.cnr.it).}
 \thanks{J. L. Sanz and D. Herranz are with the IFCA, University of Cantabria, Avda. Los Castros s/n
39005, Santander, Spain, (e-mail: sanz@ifca.unican.es;
herranz@ifca.unican.es). }
\thanks{B. Sankur is with the Bogazici University, Electrical \& Electronics Eng. Dept.,
 34342, Bebek, Istanbul, Turkey, (e-mail: bulent.sankur@boun.edu.tr).} }
%
%
%
\markboth{IEEE Transaction on Image Processing,~Vol.~?,
No.~?,~???????~201?}{Kayabol \MakeLowercase{\textit{et al.}}:
Adaptive Langevin Sampler for Separation of
\textit{t}-distribution Modelled Astrophysical Maps}
%



\maketitle

\begin{abstract}
We propose to model the image differentials of astrophysical
source maps by Student's $t$-distribution and to use them in the
Bayesian source separation method as priors. We introduce an
efficient Markov Chain Monte Carlo (MCMC) sampling scheme to unmix
the astrophysical sources and describe the derivation details. In
this scheme, we use the Langevin stochastic equation for
transitions, which enables parallel drawing of random samples from
the posterior, and reduces the computation time significantly (by
two orders of magnitude). In addition, Student's $t$-distribution
parameters are updated throughout the iterations. The results on
astrophysical source separation are assessed with two performance
criteria defined in the pixel and the frequency domains.
\end{abstract}


%
\IEEEpeerreviewmaketitle

\section{Introduction}
%
%
%
%

\PARstart{T}{he} Bayesian framework, which enables the inclusion
of prior knowledge in problem formulation, has recently been
utilized to improve the performance of Blind Source Separation
(BSS) techniques. In the context of image separation, one obvious
type of prior information is the spatial \cite{Kayabol09ip} or
spatio-chromatic \cite{Kayabol09} dependence among the source
pixels.

While there are three conditions that ensure separability of
sources, namely, non-Gaussianity, non-whiteness and
non-stationarity \cite{Cardoso01}, we choose to exploit spatial
correlation (i.e. spatial non-whiteness). The prior densities are
constituted by modeling the image differentials in different
directions as Multivariate Student's $t$-distributions
\cite{Student}. The $t$-distribution has some convenient
properties for our model: If the degree of freedom parameter of
the distribution goes to infinity, it approaches a normal density;
conversely, if the degree of freedom parameter equals 1, the
density becomes Cauchy. Therefore the $t$-distribution is a
flexible and tractable statistical model for data ranging from
broad-tailed to normally distributed. The first examples of use of
$t$-distribution in inverse imaging problems can be found in
\cite{Higdon94} and \cite{Prudyus01}. In \cite{Prudyus01}, it is
shown that the $t$-distribution approximates the distribution of
the wavelet coefficients of an image more accurately. In recent
papers, it has been used for image restoration \cite{Chantas08}
and deconvolution \cite{Tzikas09}. Notice that the degree of
freedom parameter of the $t$-distribution has the same role as the
regularization parameter of the Markov Random Field (MRF) models.
The MRF prior with Cauchy density, which was first proposed in
\cite{Hebert89} for inverse imaging, was used in
\cite{Kayabol09ip} for source separation. The model used in
\cite{Kayabol09ip} is an approximation to the $t$-distribution,
which is presented in Section \ref{sectionmojopost}.

The $t$-distribution has already been used in Bayesian audio
source separation \cite{Fevotte06} to model the discrete cosine
transform coefficients of the audio signals. It was reported that
the $t$-distribution prior had improved the sound quality over the
finite mixture-of-Gaussians prior. In this study, to solve the
Bayesian BSS problem for images without incurring in smoothing
artifacts, we propose the $t$-distribution for modeling the local
pixel differences.

We use the joint posterior density of the complete variable set to
obtain a joint estimate of all the variables. In this Bayesian
approach, the BSS problem can be solved by maximizing the joint
posterior density of the sources, the mixing matrix and the source
prior model parameters \cite{Knuth99}, \cite{Djafari99}. A method
for solving the joint posterior modal estimation problem is the
Iterated Conditional Mode (ICM) method, which maximizes the
conditional densities sequentially for each variable
\cite{Rowe99}. If the mode of the conditional density cannot be
found analytically, any deterministic optimization method can be
used \cite{Djafari99}. However, under any non-Gaussian hypothesis,
ICM does not guarantee a unique global solution.

Another algorithm suitable for learning the Gaussian MRF is the
Expectation-Maximization (EM) method. Using the Mean Field
Approximation (MFA), the expectation step of the EM algorithm can
be calculated analytically \cite{Tonazzini06}. The MFA under
Gaussian model assumption causes smoothing the edges in the image.
The reason underlying these smoothing effects is that the Gaussian
approximation violates the edge preserving property of the prior
density, and the effect is proportional to the amount of noise.
The image model with spatially varying variance parameter in
variational Bayesian approximation can help overcome the smoothing
problem \cite{Chantas08}, \cite{Tzikas09}. In \cite{Kuruoglu04},
\cite{Tonazzini06}, deterministic optimization techniques have
been used for the MRF. In \cite{Kayabol09ip}, a Gibbs sampling
stochastic optimization procedure is used. Since it is not
possible to draw samples in a simple way due to the MRF priors in
Gibbs distribution form, a Metropolis embedded Gibbs sampling has
been adopted. In \cite{Kayabol09ip}, it is reported that the Monte
Carlo results with less image smoothing, but the cost of avoiding
smoothing artifacts is a significant increase in the convergence
time.

In this study, we propose a more efficient Monte Carlo Markov
Chain (MCMC) sampling method in lieu of random walk Metropolis. To
produce proposal samples in parallel, we resort to the Langevin
stochastic equation \cite{Langevin1908}, \cite{Rossky78},
\cite{Neal93}, while the proposed samples are accepted or rejected
by the Metropolis scheme. In statistical physics, the Langevin
equation \cite{Langevin1908} is used to describe the Brownian
motion of particles in a potential field and has been used to
obtain a smart MC algorithm in \cite{Rossky78}. Another parallel
sampling algorithm is the Hamilton Monte Carlo, which is the
generalized version of the Langevin sampler \cite{Neal93}. We
conjecture that, with the samples produced in parallel by the
Langevin equation, the convergence time of the algorithm will be
significantly reduced.

Parameter estimation in Bayesian edge preserving inverse imaging
problems with Gibbs distributions is a troublesome process because
of the partition function. Although there are some methods
\cite{Higdon97,Descombes99} that calculate the parameters using
Monte Carlo techniques, their computational burdens are
prohibitive. One can resort to the Pseudo Likelihood (PL)
approximation to make the partition function separable, which is
more convenient for parameter estimation with the Maximum
Likelihood (ML) method. In \cite{Molina99}, two Bayesian
approaches have been used to estimate parameters, namely, the
Maximum-a-Posteriori (MAP) and evidence approaches. An approach to
estimate the regularization parameter from the PL approximation
has been recently proposed \cite{Kayabol09b}. The multivariate
$t$-distribution is also a PL approximation to MRF and has
advantages over MRF in parameter calculations.

There are two types of parameters in edge preserving inverse
imaging. The first one is the adaptive edge preserving parameter,
which is also known as threshold parameter. We interpret the
threshold parameter as the scale parameter of the
$t$-distribution. The other parameter is the regularization
parameter, which adjusts the balance between the likelihood and
the prior. The regularization parameter corresponds to the degree
of freedom (dof) parameter of the $t$-distribution. To estimate
the scale and dof parameters of our $t$-distribution, we use ML
estimation via EM algorithm as in \cite{Liu95}. A similar approach
has been used in \cite{Chantas08} and \cite{Tzikas09}.

In a comparative study among image source separation algorithms
\cite{Kayabol09ip}, we have found that the Bayesian formulation
with MRF prior and Gibbs sampling outperformed the heuristic and
the other Bayesian approaches. This work is based on
\cite{Kayabol09ip}, and aims to achieve a much faster MCMC
implementation without compromising its good performance. With
this goal in mind, we have been testing our algorithms on a
current problem of modern astrophysics: the separation of
radiation source maps from multichannel images of the sky at
microwave frequencies. In particular, we have been applying our
algorithm to the separation of the Cosmic Microwave Background
(CMB) radiation from the galactic emission (synchrotron and
thermal dust emissions) using realistically simulated sky maps.

The original contributions of the paper hinge on two aspects.
First, we adopt the $t$-distribution to build a prior model of the
source maps. More specifically, the $t$-distribution is used to
model the differentials of the sources, as done by the first-order
homogeneous MRF models. This is advantageous because the
flexibility of the $t$-distribution allows each source
differential to assume a different model whether impulsive or
Gaussian, simply by setting the dof parameter. The second
contribution is the introduction of the Langevin sampling scheme
in lieu of the random walk Gibbs sampling. As opposed to
pixel-by-pixel sampling, Langevin sampling generates the samples
in parallel, thus leading to much faster convergence. Furthermore,
the samples are drawn in an informed way, since their generation
follows the gradient descent direction on an energy surface.

Section \ref{BolAstro} is a brief introduction to astrophysical
sources. In Section \ref{BolBSS}, the component separation problem
in observational astrophysics is stated. Section \ref{BolBayes}
lays out the Bayesian formulation of the problem. Section
\ref{BolEst} presents the derivation steps of the adaptive
Langevin sampler algorithm along with the EM parameter estimation
method. The simulation results are presented in Section
\ref{BolSim} and interpreted in Section \ref{BolVargi}.
\section{An Introduction to Astrophysical Sources}
\label{BolAstro}
Here, we only give a brief description of the astrophysical
radiations considered, referring the interested reader to
\cite{Planck} for details. We are interested in the frequency
range 30 to 1000 GHz where the dominant diffuse radiations are the
CMB, the galactic synchrotron radiation and the thermal emission
from galactic dust. Studying this radiation would help us to
understand the distribution and the features of interstellar dust
in our galaxy.
\begin{figure}[t]
\centering{ \subfigure[CMB
horizontal]{\includegraphics[width=1.6in]{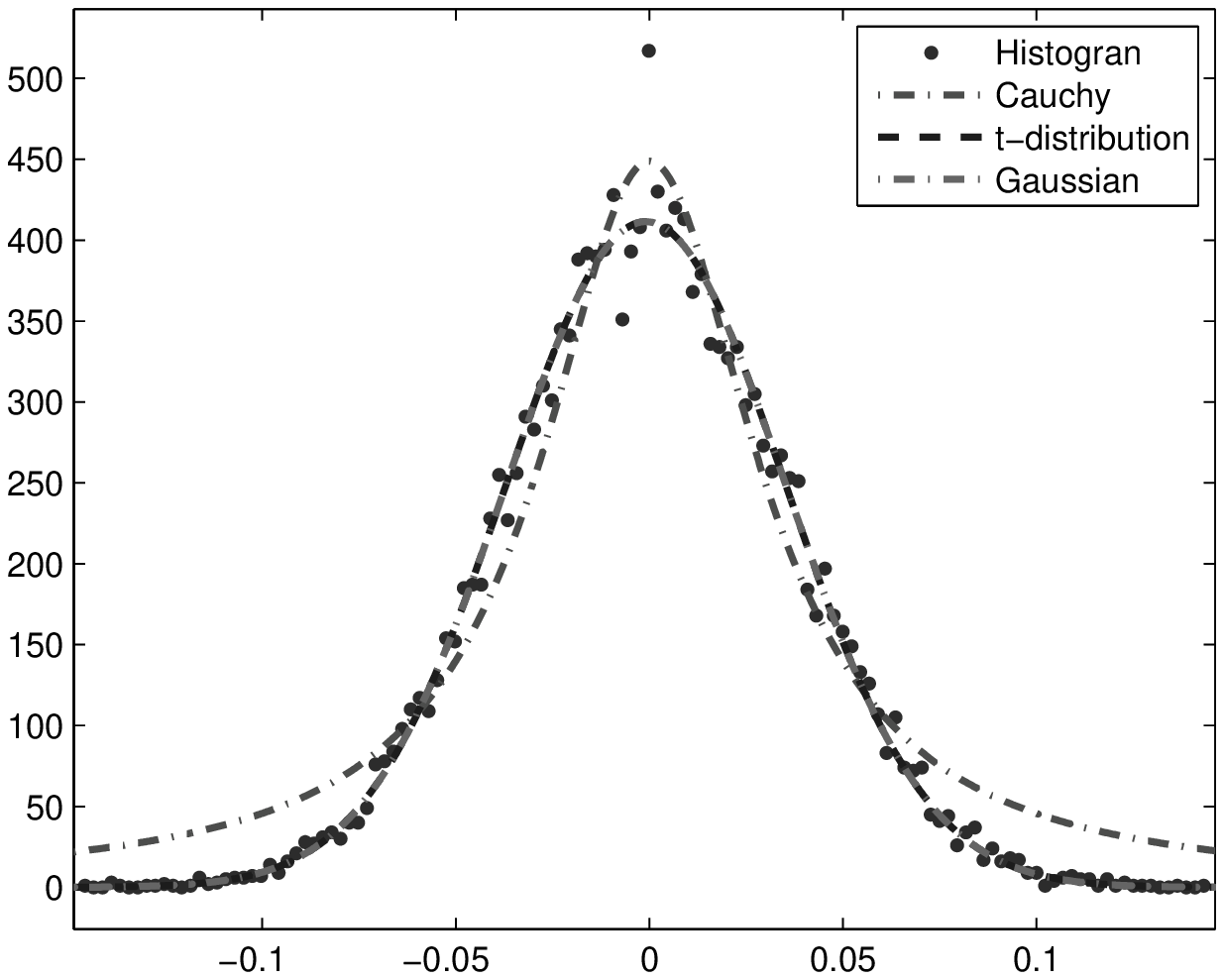}
\label{fig_first_case}} \hfil \subfigure[CMB
vertical]{\includegraphics[width=1.6in]{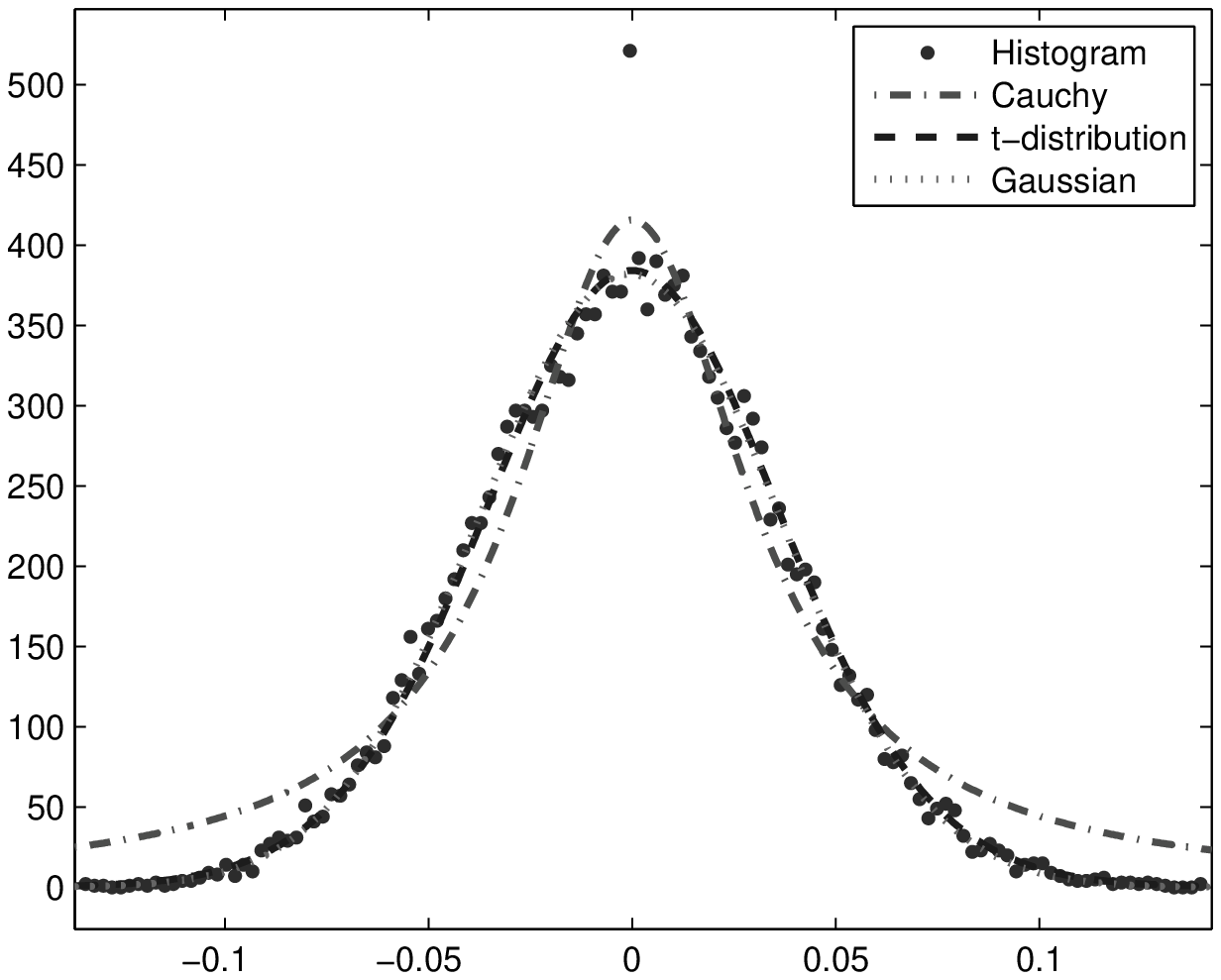}
\label{fig_first_case}}\\
\subfigure[Synchrotron
horizontal]{\includegraphics[width=1.6in]{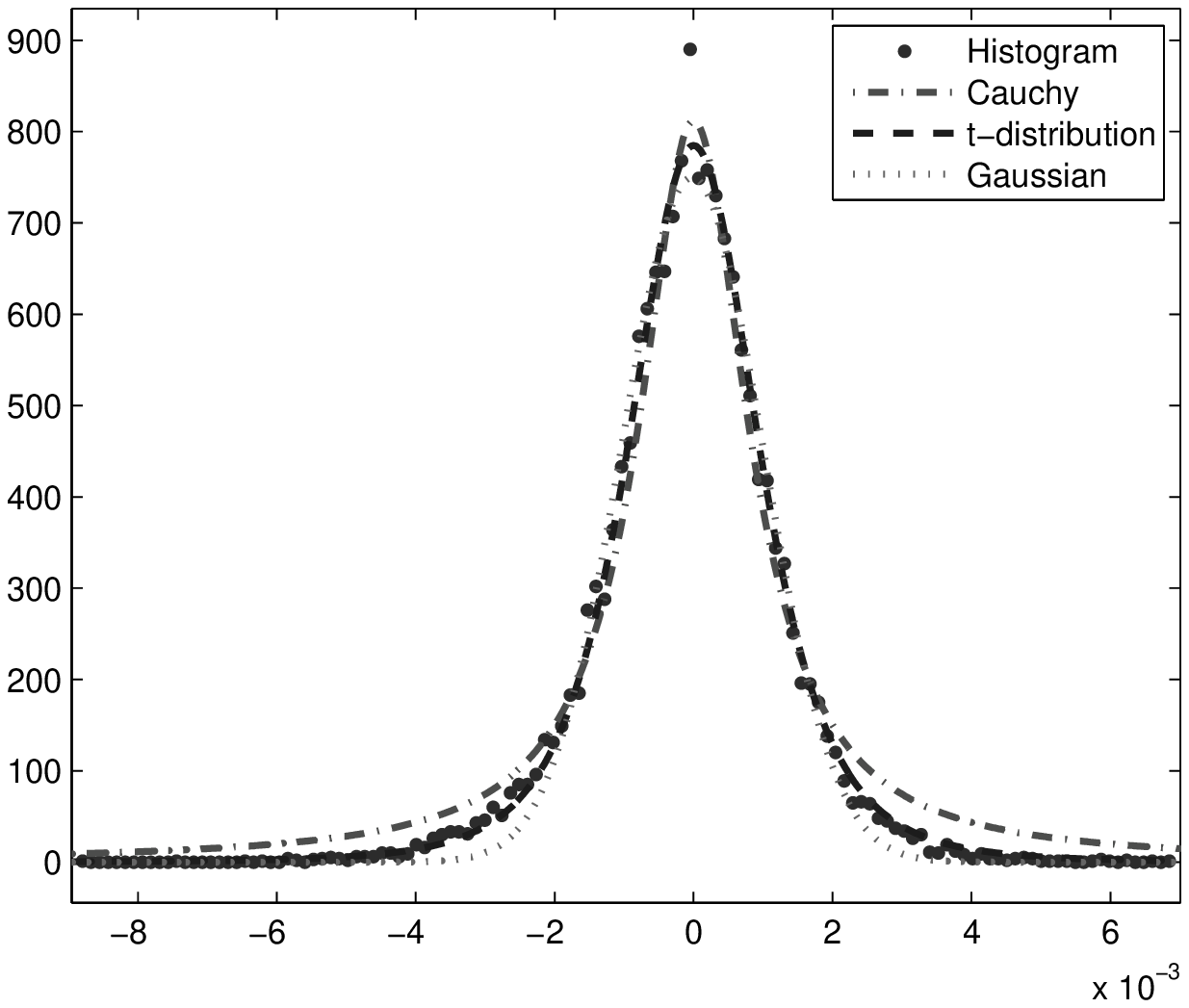}
\label{fig_second_case}} \hfil \subfigure[Synchrotron
vertical]{\includegraphics[width=1.6in]{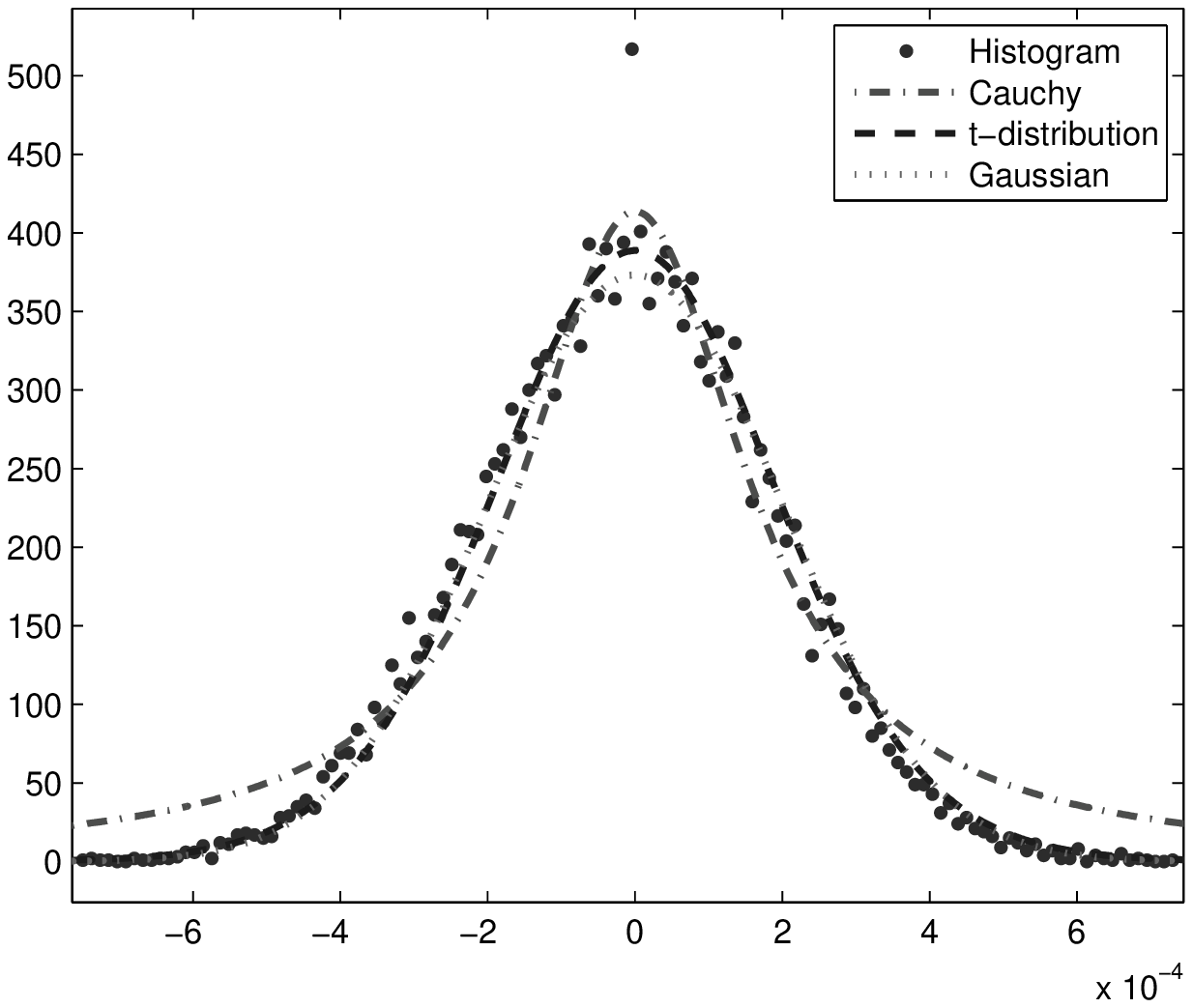}
\label{fig_second_case}} \\ \subfigure[Dust
horizontal]{\includegraphics[width=1.6in]{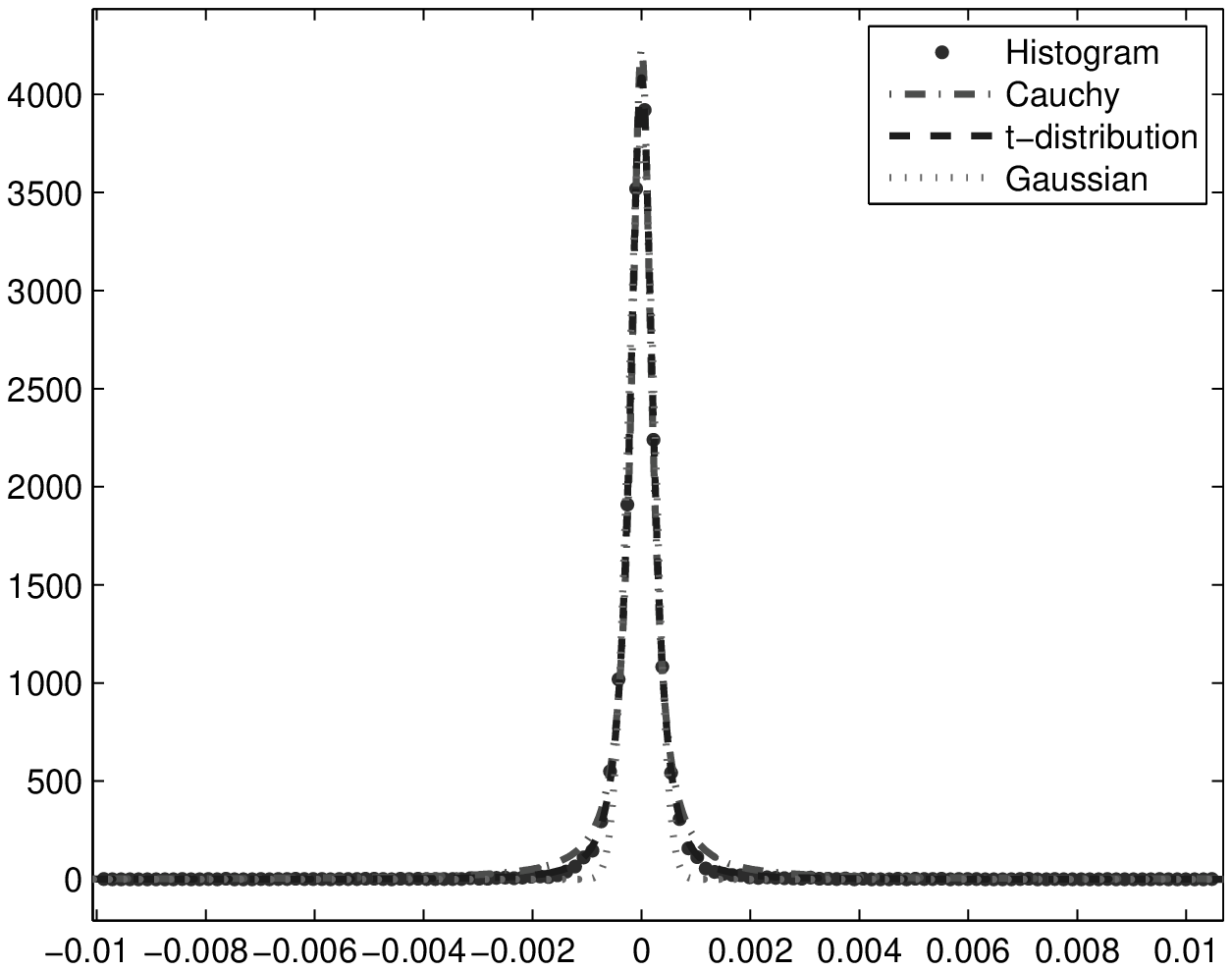}
\label{fig_second_case}} \subfigure[Dust
vertical]{\includegraphics[width=1.6in]{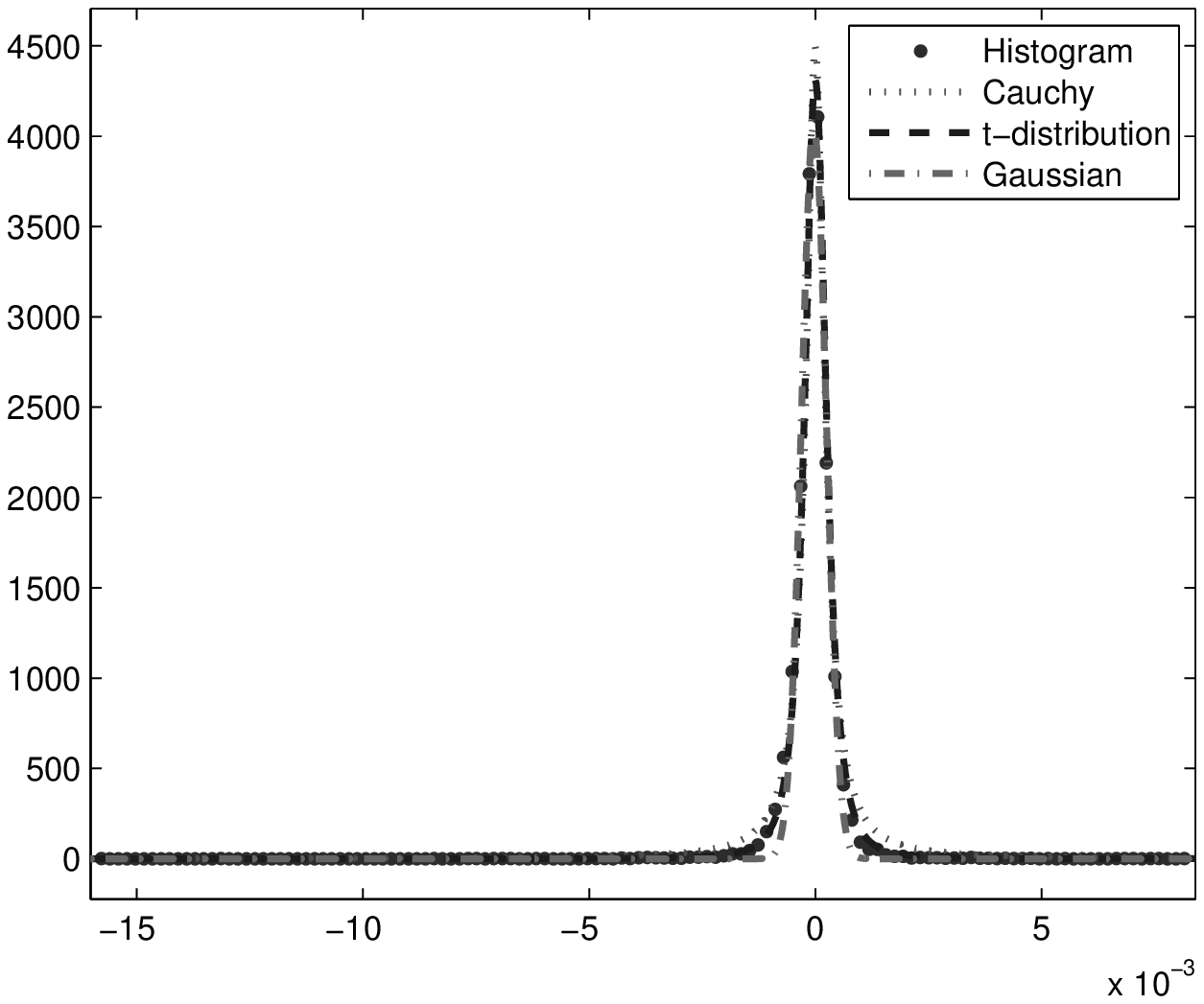}
\label{fig_second_case}}}%
\caption{Fitting plots of the image differential histograms (dots)
of CMB, synchrotron and dust components in horizontal and vertical
directions. The fitted functions proportional to Gaussian (dot
line), Cauchy (dash-dot line) and $t$-distribution (dash line).}
\label{Ffit}
\end{figure}

The most interesting astrophysical source in the microwave region
of the electromagnetic spectrum is the CMB, a relic radiation
originating from the time when the Universe was ~300.000 years
old. The discovery of the CMB is one of the fundamental milestones
of modern cosmology and its study allows us to determine
fundamental parameters such as the age of the universe, its matter
and energy composition, its geometry and many other relevant
cosmological parameters. CMB should be a blackbody radiation at a
temperature of $2.726$ K, thus its emission spectrum should be
perfectly known. The CMB emission dominates over the other sources
at frequencies around 100 GHz. The CMB temperature is not
perfectly anisotropic. Standard cosmological models predict that
the CMB anisotropy is Gaussian distributed, although some
alternative models permit a certain degree of non-Gaussianity.
Current observations are compatible with the hypothesis of
Gaussianity. Two all-sky surveys have been made on CMB so far, by
NASA's satellites COBE \cite{COBEweb} and WMAP \cite{WMAPweb}. A
European mission whose data will be highly accurate and spatially
resolved, Planck \cite{Planck}, is about to provide its first
full-sky coverage maps.

The CMB signal is mixed with other astrophysical sources of
electromagnetic radiation. Relativistic electrons being
accelerated by magnetic fields in the Galaxy give rise to
synchrotron emission, which dominates over the CMB in regions
close to the Galactic plane especially at low frequencies ($< 200$
GHz). According to observations in other frequency bands, the
synchrotron emission spectrum follows a power law with a negative
exponent, whose value is presently known with high uncertainty.
Synchrotron is the dominant radiation in the low-frequency bands
of our range of interest. Inter-stellar dust grains are heated by
nearby stars and re-emit thermal radiation in the far infrared
region of the electromagnetic spectrum. Dust radiation is dominant
in the high end of our range. In particular, it is almost the only
significant contribution to the total diffuse radiation between
800 GHz and 1000 GHz. Its emission spectrum should follow a
greybody law, with unknown spectral index and an additional degree
of freedom given by the thermodynamical temperature of the dust
grains. For a short review on CMB astronomy, see \cite{Hu02}.

There are other astrophysical sources present at microwave
frequencies, such as free-free emission due to free electrons,
anomalous dust emission and radiation coming from extragalactic
sources, but their relevance is smaller. In this work we will
focus on the main three astrophysical sources present in CMB
experiments: CMB, synchrotron and dust.

In order to justify our adoption of the Student's $t$-distribution
in a relevant case, we have selected a $15^{\circ}\times
15^{\circ}$ sky patch, located at $0^{\circ}$ galactic longitude
and $40^{\circ}$ galactic latitude, discretized into a $512\times
512$-pixel map. Within this patch, we have introduced simulated
CMB, synchrotron, and dust radiation maps (as in
\cite{Bonaldi06}). We have computed the source image differentials
for horizontal and vertical directions, and estimated their
empirical distributions. We have fitted three different functions
to the empirical distributions of the image differentials of the
astrophysical sources with nonlinear least square method using the
Curve Fitting Toolbox of MATLAB. These functions are proportional
to Gaussian, Cauchy and $t$-distribution. Fig. \ref{Ffit} shows
the fitting results for CMB, synchrotron and dust images. Table
\ref{Tfit} lists the residual Root Mean Square Errors (RMSE) of
the fits. The Gaussian gives the best fit for CMB because the CMB
is theoretically distributed as a Gaussian \cite{Hu02}. Overall,
the $t$-distribution appears to be a good choice for modeling the
image differential statistics in horizontal and vertical
directions of all the components. The estimated dof parameters of
$t$-distributions show that indeed the proposed model assumes from
impulsive to Gaussian characteristic underlying each component. If
the component is Gaussian as CMB, the dof parameter becomes bigger
and if it is impulsive, the dof parameter becomes very small.
\begin{table}
  \caption{Root Mean Square Error (RMSE) of fitting of the image differential
histograms of CMB, synchrotron and dust components. The last
column shows the estimated dof parameters of the
$t$-distribution.}
  \label{Tfit}
  \centering
\begin{tabular}{|c||c|c|c|c|}
  \multicolumn{1}{}{} &   \multicolumn{4}{c}{Horizontal direction}     \\
   \hline
   & Gaussian \hspace{6pt}& Cauchy \hspace{8pt} & $t$-distribution & dof  \\
  \hline
   CMB              & 15.11 & 30.84 & 15.47 & 25.71\\
   Synchrotron      & 23.70 & 30.06 & 15.70 & 3.59\\
   Dust             & 67.99 & 33.21 & 13.84 & 1.81\\
  \hline
\end{tabular}
\begin{tabular}{|c||c|c|c|c|}
  \multicolumn{1}{}{} &   \multicolumn{4}{c}{Vertical direction}     \\
   \hline
   & Gaussian \hspace{6pt}& Cauchy \hspace{8pt} & $t$-distribution & dof  \\
  \hline
   CMB              & 15.75 & 29.70 & 15.83 & 15.75\\
   Synchrotron      & 19.11 & 31.00 & 18.39 & 19.11\\
   Dust             & 63.73 & 68.19 & 54.42 & 2.18\\
  \hline
\end{tabular}
\end{table}
\section{Component Separation Problem in Observational Astrophysics}
\label{BolBSS}
Virtually any application in observational astrophysics has to do
with problems of component separation. Indeed, all the
astrophysical observations result from the superposition of the
radiation sources placed along the line of sight. While very
distant sources can be distinguished by the redshift analysis, for
nearby sources this is not possible. In fact, physically distinct
sources can sometimes be found within a close range of each other.
Furthermore, high sensitivity and high resolution measurements can
give rise to source mixing problems even in the cases where the
radiation under study is dominant over interfering radiations.
Apart from redshift analysis, useful methods to distinguish
between superimposed physically different radiations include
spectral analysis and morphological analysis. In this paper, we
only treat the former approach, exploiting the differences in the
emission spectra shown by physically distinct radiation sources.
This implies that the separation must be done on the basis of
measurements made at different frequency bands.

We assume that the observed images, $y_{k}, k \in
\{1,2,\ldots,K\}$, are linear combinations of $L$ source images.
Let the $k$th observed image be denoted by $y_{k,i}$, where $i \in
\{ 1,2,\ldots,N \}$  represents the lexicographically ordered
pixel index. The image separation problem consists in finding $L$
independent sources from $K$ different observations. If
$\mathbf{s}_{l}$ and $\mathbf{y}_{k}$ denote $N\times 1$ vector
representations of source and observation images, respectively,
then the observation model can be written as
\begin{equation}\label{mixeniyi}
    \mathbf{y}_{k}=\sum_{l=1}^{L}a_{k,l}\mathbf{s}_{l}+\mathbf{n}_{k},
    \qquad k=1,\ldots,K
\end{equation}
where $\mathbf{n}_{k}$ is an iid zero-mean noise vector with
$\Sigma = \sigma_{k}^{2} \mathbf{I}_{N}$ covariance matrix and
$\mathbf{I}_{N}$ is an identity matrix. Although the noise is not
necessarily homogeneous in the astrophysical maps, in this study
we assume that the noise variance is homogeneous within each sky
patch and is also known.

Since the observation noise is assumed to be independent and
identically distributed zero-mean Gaussian at each pixel, the
likelihood is expressed as
\begin{eqnarray}\label{l}
    p(\mathbf{y}_{1:K} | \mathbf{s}_{1:L}, \mathbf{A}) &\propto& \prod_{k=1}^{K}  \exp  \left\{ - W(\mathbf{s}_{1:L}|\mathbf{y}_{k},\mathbf{A},\sigma_{k}^{2}) \right\} \\
    \label{l2} W(\mathbf{s}_{1:L}|\mathbf{y}_{k},\mathbf{A},\sigma_{k}^{2}) &=&
    \frac{||(\mathbf{y}_{k}-\sum_{l=1}^{L}a_{k,l}\mathbf{s}_{l})||^2}{2\sigma_{k}^{2}}
\end{eqnarray}
where the mixing matrix $\mathbf{A}$ contains all the mixing
coefficients $a_{k,l}$ introduced in (\ref{mixeniyi}).

For many purposes, a mixing model of the type (\ref{mixeniyi}) is
considered to fit reasonably well to an astrophysical observation.
The details on how to get an equation similar to (\ref{mixeniyi})
from the physics of the problem can be found in
\cite{Baccigalupi00}. Here, we only summarize the main assumptions
made with this purpose. First, we assume that the superposition of
the signals originating from different sources is linear and
instantaneous. In the astrophysical case, this assumption is
clear, since the physical quantities to be measured are
superpositions of electromagnetic waves coming, for any bearing,
exactly from the same line of sight without any scattering or
diffraction effect. The second assumption in modeling
astrophysical observations is that each source has an emission
spectrum that does not vary with the bearing. This assumption
implies that individual radiations result from the product of a
fixed spatial template and an isotropic emission spectrum. Both
assumptions need closer attention. The precise emission spectrum
generated by any physical phenomenon depends on many quantities
that may not all be distributed uniformly in the sky. Although in
many applications the isotropy assumption has been adopted
successfully, in many other cases the space-variability of the
radiation sources must be taken into account to allow a good
separation to be performed. Furthermore, if the effect of the
telescope is taken into account, then the instantaneous model is
no more valid since, for the finite aperture, the light captured
in a fixed direction in the telescope does not come from that
direction alone. In formulas, model (\ref{mixeniyi}) becomes
\begin{equation}\label{mix2}
    \tilde{\mathbf{y}}_{k}=\mathbf{h}_{k}\ast \mathbf{y}_{k} = \mathbf{h}_{k}\ast\sum_{l=1}^{L}a_{k,l}\mathbf{s}_{l} + \mathbf{n}_{k}
\end{equation}
where the asterisk means convolution, and $\mathbf{h}_{k}$ is the
telescope radiation pattern in the $k$'th observation channel.
Note that, if $\mathbf{h}_{k}$ is the same for all the channels,
(\ref{mix2}) can be written as
\begin{equation}\label{mix3}
    \tilde{\mathbf{y}}_{k} = \sum_{l=1}^{L}a_{k,l}\mathbf{h}\ast
    \mathbf{s}_{l}+\mathbf{n}_{k}= \sum_{l=1}^{L}a_{k,l}\tilde{\mathbf{s}}_{l} + \mathbf{n}_{k}
\end{equation}
and the problem is again instantaneous for the modified sources
$\tilde{\mathbf{s}}_{l}$, which are the physical sources smoothed
by the common radiation pattern $\mathbf{h}$. Unfortunately,
especially in the radio- to millimeter-wave ranges, the telescope
aperture depends strongly on frequency, and model (\ref{mix3})
cannot be adopted directly, unless the observed signals are
preprocessed to reduce their angular resolution to the worst
available (see \cite{Bonaldi06}).

Since no imaging system can achieve an infinite resolution, we
should use (\ref{mix2}) as our generative model. However, to avoid
problems with the convolutive mixtures, we assume to have a
telescope with the same radiation pattern in all the channels, so
as to be able to use model (\ref{mix3}). Note that this can always
be obtained by preprocessing, provided that all the beam patterns
are known. Hereafter, as it will not cause any ambiguity, we drop
the tilde accent from the symbols used to denote the data and the
source vectors.
\section{Source Separation Defined in the Bayesian
Framework} \label{BolBayes}
\subsection{Source Model}
\begin{figure}[]
 \centerline{\includegraphics[width=3.8in]{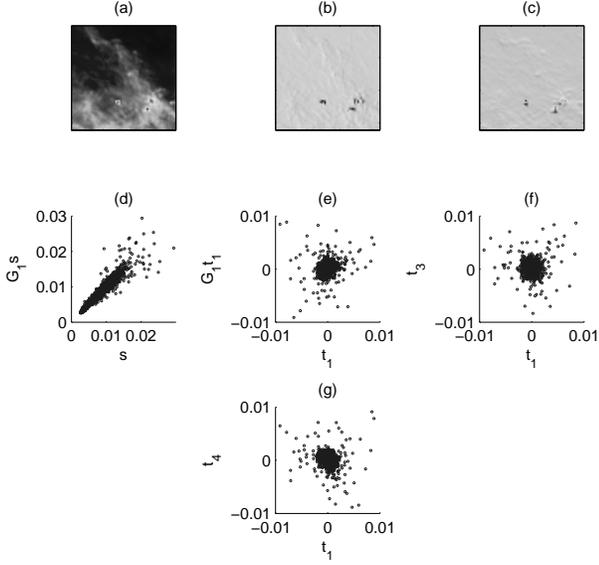}}
\caption{(a): The synchrotron map $\mathbf{s}$, (b): Right
$\mathbf{t}_{1} = \mathbf{s} - \mathbf{G}_{1}\mathbf{s}$ and (c):
Up $\mathbf{t}_{3} = \mathbf{s} - \mathbf{G}_{3}\mathbf{s}$
difference maps of (a). (d): Scatter-plot of the first order right
neighbor pixels of (a), $\mathbf{s}$ and
$\mathbf{G}_{1}\mathbf{s}$. (e): Scatter-plot of the first order
right neighbor pixels of (b), $\mathbf{t}_{1}$ and
$\mathbf{G}_{1}\mathbf{t}_{1}$. (f): Scatter-plot of the right
difference (b) versus up difference (c), $\mathbf{t}_{1}$ and
$\mathbf{t}_{3}$. (g): Scatter-plot of the right difference (b)
versus left difference (c), $\mathbf{t}_{1}$ and $\mathbf{t}_{4}$.
}
 \label{autodependust}
\end{figure}

Neighbor pixels in our images have strong dependency. This is
demonstrated, for example, in Fig. \ref{autodependust}(d) where
the scatter-plot shows the first order right neighbor pixels of
the dust map shown in Fig. \ref{autodependust}(a). The dependency
decreases in the high intensity region. This region in the
scatter-plot corresponds to spatially localized structures with
high image intensity of the map in Fig. \ref{autodependust}(a).
The existence of a small number of point-like structures in the
maps indicates that the dependency assumption is valid. In view of
this, we can write an auto-regressive source model using the first
order neighbors of the pixel:
\begin{equation}\label{AR}
    \mathbf{s}_{l} = \alpha_{l,d}\mathbf{G}_{d}\mathbf{s}_{l} + \mathbf{t}_{l,d}
\end{equation}
where $d \in \{1,\ldots, D\}$ denotes one of the main directions
(left, right, up and down) and $D=4$ is the cardinality of the set
of image differential directions. Matrix $\mathbf{G}_{d}$ is a
linear one-pixel shift operator in direction $d$, $\alpha_{l,d}$
is the regression coefficient and the regression error
$\mathbf{t}_{l,d}$ is an iid $t$-distributed zero-mean vector with
dof parameter $\beta_{l,d}$ and scale parameters $\delta_{l,d}$,
$\mathcal{T}(\mathbf{t}_{l,d}|0,\delta_{l,d}\mathbf{I}_{N},\beta_{l,d})$.
We can justify the iid assumption of $\mathbf{t}_{l,d}$ by
plotting (see Fig. \ref{autodependust}(e)) the values in
$\mathbf{t}_{l,d}$ versus its first order neighbors,
$\mathbf{G}_{d}\mathbf{t}_{l,d}$. We can interpret
$\mathbf{t}_{l,d}$ as a decorrelated version of $\mathbf{s}_{l}$.
Fig. \ref{autodependust}(b) shows $\mathbf{t}_{l,d}$ for $d=1$. By
comparing Fig. \ref{autodependust}(d) and (e), we can say that
$\mathbf{t}_{l,d}$ is spatially more independent than
$\mathbf{s}_{l}$.

If the image $\mathbf{s}_{l}$ were Gaussian distributed, then the
regression error would also be Gaussian. However in real images,
the regression error is better modelled by some heavy-tailed
distribution. The $t$-distribution can conveniently model the
statistics of data whose distribution ranges from Cauchy to
Gaussian, and therefore it is a convenient model for the
statistics of the high spatial frequency contents of images, such
as regression errors \cite{Prudyus01}. The scale parameter
$\delta_{l,d}$ can be assumed as a space-varying parameter to
model the highly non-stationary sources, i.e. sparse sources, but
homogenous variance assumption has been observed to be adequate
for diffuse astrophysical source images for the image patch sizes
that we use in the simulations. We use a homogeneous variance
since, in our case, the increased complexity derived from
inhomogeneity is not justified by a significant improvement in
performance.

The regression error $\mathbf{t}_{l,d}$ represents the directional
image differential in the direction $d$. The multivariate
probability density function of an image modelled by a
$t$-distribution can be defined as
\begin{eqnarray}\label{pT}\nonumber
    p(\mathbf{t}_{l,d}|\alpha_{l,d},\beta_{l,d},\delta_{l,d}) & = &
    \frac{\Gamma((N+\beta_{l,d})/2)}{\Gamma(\beta_{l,d}/2)(\pi\beta_{l,d}\delta_{l,d})^{N/2}}\\
    & & \times \left[1+\frac{\phi_{d}(\mathbf{s}_{l},\alpha_{l,d})}{\beta_{l,d}\delta_{l,d}}\right]^{-(N+\beta_{l,d})/2}
\end{eqnarray}
where
$\phi_{d}(\mathbf{s}_{l},\alpha_{l,d})=||\mathbf{t}_{l,d}||^2 =
||\mathbf{s}_{l} - \alpha_{l,d}\mathbf{G}_{d}\mathbf{s}_{l}||^2$
and $\Gamma(.)$ is the Gamma function. We can write the density of
$\mathbf{s}_{l}$ by using the image differentials in different
directions, assuming directional independence, as
$p(\mathbf{s}_{l}|\Theta) = \prod_{d=1}^{D}
p(\mathbf{t}_{l,d}|\alpha_{l,d},\beta_{l,d},\delta_{l,d})$ where
$\Theta = \{ \alpha_{1:L,1:D}, \beta_{1:L,1:D},
\delta_{1:L,1:D}\}$. We can simply justify this assumption by
plotting the horizontal regression error versus the vertical one
as shown in Fig. \ref{autodependust}(f). We can observe in this
figure that the scatter plot of the regression errors in left and
up directions is almost circular and this justifies our assumption
of independence. Fig. \ref{autodependust}(g) shows the
scatter-plot of right difference versus left difference. In spite
of a weak dependence between right and left difference images, we
maintain the independency assumption to constitute the products of
the regression error probabilities model. A similar approach to
constitute a single prior by multiplying different individual
priors can be found in \cite{Hinton99}. In this study, we use the
left/right and up/down differences jointly in the prior model to
balance their contributions. Otherwise, we could obtain
directionally biased results.

The $t$-distribution can be written in implicit form by using a
Gaussian and a Gamma density. The $t$-distribution has a dof
parameter $\beta$, which is itself governed by Gamma distribution
with parameter $\beta/2$. The $t$-distribution has the following
form \cite{Liu95}:
\begin{eqnarray}\label{jointT}\nonumber
      p(\mathbf{t}_{l,d}|\alpha_{l,d},\beta_{l,d},\delta_{l,d})  =  \int
    p(\mathbf{t}_{l,d}|\nu_{l,d},\delta_{l,d})p(\nu_{l,d}|\beta_{l,d}) d\nu_{l,d} \\
        =  \int
    \mathcal{N}\left(\mathbf{t}_{l,d}|0,\frac{\delta_{l,d}\mathbf{I}_{N}}{\nu_{l,d}}\right)
    \mathcal{G}\left(\nu_{l,d}|\frac{\beta_{l,d}}{2},\frac{\beta_{l,d}}{2}\right)d\nu_{l,d}.
\end{eqnarray}
The interpretation of this equation is that if
$\mathbf{t}_{l,d}|\delta_{l,d},\nu_{l,d}$ is distributed with a
normal density
$\mathcal{N}(\mathbf{t}_{l,d}|0,\delta_{l,d}\mathbf{I}_{N}/\nu_{l,d})$
and the parameter $\nu_{l,d}$ has a Gamma prior as
$\mathcal{G}(\nu_{l,d}|\beta_{l,d}/2,\beta_{l,d}/2)$, the
distribution of $\mathbf{t}_{l,d}|\beta_{l,d},\delta_{l,d}$
becomes a $t$-distribution such that
$\mathcal{T}(\mathbf{t}_{l,d}|0,\delta_{l,d}\mathbf{I}_{N},\beta_{l,d})$.
The representation in (\ref{jointT}) is a particular case of the
Gaussian Scale Mixture (GSM) densities. The ML estimation of the
parameters $\alpha_{l,d}$, $\beta_{l,d}$ and $\delta_{l,d}$ using
the EM method \cite{Liu95} is given in Section \ref{BolReg}.
\subsection{Posteriors}\label{sectionjopost}
The joint posterior density of all the unknowns in the BSS problem
can be written as:
\begin{equation}\label{bayesfull}
    p(\mathbf{s}_{1:L}, \mathbf{A},\Theta|\mathbf{y}_{1:K}) \propto
     p(\mathbf{y}_{1:K}|\mathbf{s}_{1:L}, \mathbf{A}) p(\mathbf{s}_{1:L}, \mathbf{A},\Theta)
\end{equation}
where $p(\mathbf{y}_{1:K} | \mathbf{s}_{1:L}, \mathbf{A})$ is the
likelihood and $p(\mathbf{s}_{1:L}, \mathbf{A},\Theta)$ is the
joint prior density of unknowns. The joint prior can be factorized
as $p(\mathbf{s}_{1:L}|\alpha_{1:L,1:D}, \beta_{1:L,1:D},
\delta_{1:L,1:D})$ $p(\mathbf{A})$ $p(\beta_{1:L,1:D})$
$p(\delta_{1:L,1:D})$ $p(\alpha_{1:L,1:D})$. Furthermore, since
the sources are assumed to be independent, the joint probability
density of the sources is also factorized as
$p(\mathbf{s}_{1:L}|\Theta) =
\prod_{l=1}^{L}p(\mathbf{s}_{l}|\Theta)$.

Mathematically, we can assume uniform priors for $\alpha_{l,d} \in
(-1, 1)$, $\delta_{l,d} \in (0,\infty)$ and $a_{k,l} \in
(0,\infty)$, because $a_{k,l}$'s are always positive. The
practical usage of these priors is explained in Section
\ref{BolAlg}. We use a conjugate Gamma prior for $\beta_{l,d}$
$\sim$ $\mathcal{G}(1/2,2\times 10^{-3})$. We have determined the
respective parameters experimentally. The conditional posteriors
of all model parameters are written as
\begin{eqnarray}\label{post}
\nonumber
  p(a_{k,l}|\mathbf{y}_{1:K},\mathbf{s}_{1:L}, \mathbf{A}_{-a_{k,l}},\Theta) & \propto  & p(\mathbf{y}_{1:K}|\mathbf{s}_{1:L},\mathbf{A})\\ \nonumber
  p(\alpha_{l,d}|\mathbf{y}_{1:K},\mathbf{s}_{1:L}, \mathbf{A},\Theta_{-\alpha_{l,d}}) & \propto & p(\mathbf{t}_{l,d}|\Theta) \\
  p(\beta_{l,d}|\mathbf{y}_{1:K},\mathbf{s}_{1:L}, \mathbf{A},\Theta_{-\beta_{l,d}}) & \propto & p(\mathbf{t}_{l,d}|\Theta)p(\beta_{l,d}) \\\nonumber
  p(\delta_{l,d}|\mathbf{y}_{1:K},\mathbf{s}_{1:L}, \mathbf{A},\Theta_{-\delta_{l,d}})  & \propto &  p(\mathbf{t}_{l,d}|\Theta)p(\delta_{l,d})
  \\ \nonumber
   p(\mathbf{s}_{l}|\mathbf{y}_{1:K},\mathbf{s}_{(1:L)-l}, \mathbf{A},\Theta)  & \propto & p(\mathbf{y}_{1:K}|\mathbf{s}_{1:L},\mathbf{A}) p(\mathbf{s}_{l}|\Theta)
\end{eqnarray}
where $-variable$ expressions in the subscripts denote the removal
of that variable from the variable set. The parameters $\alpha$,
$\beta$ and $\delta$ have size $L\times D$, $\mathbf{A}$ has size
$K\times L$ and the sources have size $L\times N$. Overall there
are $(3D + K + N)L$ unknowns.

For parameters $\alpha_{l,d}$, $\delta_{l,d}$ and $\beta_{l,d}$,
we exploit the EM method. To estimate the source images, we use a
version of the posterior $p(\mathbf{s}_{l}|.)$ augmented by
auxiliary variables and find the estimation with a Langevin
sampler. The details are given in Section \ref{BolEst}.
\section{Estimation of Sources and Parameters} \label{BolEst}
In this section, we give the details of the estimation of the
sources and the parameters.
\subsection{Sources} \label{sectionmojopost}
We modify the posterior densities of the source images
$p(\mathbf{s}_{l}|\Theta)$ to obtain a more efficient MCMC
sampler. In the classical MCMC schemes, a random walk process is
used to produce the proposal samples. Although random walk is
simple, it affects adversely the convergence time. The random walk
process only uses the previous sample for producing a new
proposal. Instead of a random walk, we use the Langevin stochastic
equation, which exploits the gradient information of the energy
function to produce a new proposal. Since the gradient directs the
proposed samples towards the mode, the final sample set comes
mostly from around the mode of the posterior \cite{Grenander94},
\cite{Dostert06}.

The Langevin equation can be obtained from the total energy
function. We first define everything in continuous time to give
the derivation steps of the Langevin equation, then we transfer
them into discrete time. To obtain the total energy function, we
introduce a velocity parameter $\mathbf{v}_{l}(t)=d
\mathbf{s}_{l}(t)/dt$ to define the kinetic energy such that
\begin{equation}\label{Kintx}
    K(\mathbf{v}_{l}(t)|\mathbf{M}_{l}) = \frac{1}{2} \mathbf{v}_{l}^{T}(t)\mathbf{M}_{l}\mathbf{v}_{l}(t)
\end{equation}
where $\mathbf{M}$ is a diagonal matrix whose diagonal elements
correspond to mass parameters $m_{l,n}$ for pixel index
$n=1,\ldots, N$. Using the velocity parameter $\mathbf{v}_{l}$,
the modified version of the posterior density in (\ref{post}) is
written as
$p(\mathbf{s}_{l},\mathbf{v}_{l}|\mathbf{y}_{1:K},\mathbf{s}_{(1:L)-l},
\mathbf{A},\Theta,\mathbf{M}_{l})$ $\propto$
$p(\mathbf{y}_{1:K}|\mathbf{s}_{1:L},\mathbf{A})$
$p(\mathbf{s}_{l}|\Theta)p(\mathbf{v}_{l}|\mathbf{M}_{l})$. More
explicitly, it can be written as
\begin{eqnarray}\label{augpost}\nonumber
    p(\mathbf{s}_{l},\mathbf{v}_{l}|\mathbf{y}_{1:K},\mathbf{s}_{(1:L)-l},
\mathbf{A},\Theta,\mathbf{M})  \propto  \\ \exp \{
-(W(\mathbf{s}_{1:L}|\mathbf{A}) + U(\mathbf{s}_{l}|\Theta) +
K(\mathbf{v}_{l}|\mathbf{M})) \}
\end{eqnarray}
where the energy function $U(\mathbf{s}_{l}|\Theta)$ of a source
image can be written in terms of image differentials
$\mathbf{t}_{l,d}$ as
\begin{equation}\label{Upl}
    U(\mathbf{s}_{l}|\Theta) =
    \sum_{d=1}^{D}\rho(\mathbf{t}_{l,d}|\Theta).
\end{equation}
where the function $\rho(\mathbf{t}_{l,d}|\Theta)$ is proportional
to the negative logarithm of the $t$-distribution in (\ref{pT}),
that is,
\begin{equation}\label{multicauchy}
    \rho(\mathbf{t}_{l,d}|\Theta) = \frac{N+\beta_{l,d}}{2}\log\left[1+\frac{\phi_{d}(\mathbf{s}_{l},\alpha_{l,d})}{\beta_{l,d}\delta_{l,d}}\right]
\end{equation}
and the function $\log [1+\phi_{d}(\mathbf{s}_{l},\alpha_{l,d})/
\beta_{l,d}\delta_{l,d} ]$ is the regularization function proposed
in \cite{Hebert89}. The terms $(N+\beta_{l,d})/2$ and
$\beta_{l,d}\delta_{l,d}$ correspond to the regularization and the
threshold parameters, respectively, used in edge preserving image
reconstruction.

The energy function $W(\mathbf{s}_{1:L}|\mathbf{A})$ was defined
in (\ref{l2}). The total energy function is proportional to the
negative logarithm of the posterior. In summary, the three terms
correspond, respectively, to the fit to data and to the inertial
and the kinetic energy terms. We can define the Lagrangian
function: $L(\mathbf{s}_{l}(t),\mathbf{v}_{l}(t)) =
K(\mathbf{v}_{l}) - W(\mathbf{s}_{1:L}) - U(\mathbf{s}_{l})$ and
write the Lagrange-Euler equation for the Lagrangian as follows
\begin{equation}\label{LEcont}
    \frac{d}{dt}\left( \frac{\partial L(\mathbf{s}_{l}(t),\mathbf{v}_{l}(t))}{\partial
    \mathbf{v}_{l}}\right) = \frac{\partial L(\mathbf{s}_{l}(t),\mathbf{v}_{l}(t))}{\partial
    \mathbf{s}_{l}},
\end{equation}
\begin{equation}\label{LEcont2}
   \mathbf{M}_{l}\frac{d\mathbf{v}_{l}}{dt} = - \frac{\partial }{\partial
    \mathbf{s}_{l}} E(\mathbf{s}_{l}).
\end{equation}
where $E(\mathbf{s}_{1:L})=W(\mathbf{s}_{1:L})+U(\mathbf{s}_{l})$.
If we discretize the dynamics in (\ref{LEcont2}) and velocity
$\mathbf{v}_{l}(t)$ using the Leapfrog method \cite{Neal93}, we
obtain the following three-step iteration
\begin{eqnarray}\label{LEdisc1}
    \mathbf{v}_{l}^{k+\frac{1}{2}} & = & \mathbf{v}_{l}^{k} - \frac{1}{2}
    \tau_{l} \mathbf{M}_{l}^{-\frac{1}{2}}
    \mathbf{g}(\mathbf{s}_{1:L}^{k}) \\ \label{LEdisc2}
    \mathbf{s}_{l}^{k+1} & = & \mathbf{s}_{l}^{k} +
    \tau_{l} \mathbf{v}_{l}^{k+\frac{1}{2}}\\
    \mathbf{v}_{l}^{k+1} & = & \mathbf{v}_{l}^{k+\frac{1}{2}} - \frac{1}{2}
    \tau_{l} \mathbf{M}_{l}^{-\frac{1}{2}}
    \mathbf{g}(\mathbf{s}_{1:L}^{k})
\end{eqnarray}
where $\mathbf{g}(\mathbf{s}_{1:L}^{k}) = [\nabla_{\mathbf{s}_{l}}
E(\mathbf{s}_{1:L})]_{\mathbf{s}_{1:L} = \mathbf{s}_{1:L}^{k}}$,
$\nabla_{\mathbf{s}_{l}}$ is the gradient with respect to
$\mathbf{s}_{l}$ and $\tau_{l}$ is the discrete time step. If we
define a diagonal matrix $\mathbf{D}_{l}^{\frac{1}{2}} =
\tau_{l}\mathbf{M}_{l}^{-\frac{1}{2}}$, so that, for the $n$th
pixel, the diffusion coefficient is
$\mathbf{D}_{l}(n,n)=\tau_{l}^{2}/m_{l,n}$. Matrix $\mathbf{D}$ is
referred to here as the diffusion matrix, and is derived in
Section \ref{BolDif}. Instead of this step scheme, we use the
one-step Langevin difference equation. To obtain the single step
Langevin update equation for $\mathbf{s}_{l}$, we substitute
(\ref{LEdisc1}) into (\ref{LEdisc2}).
\begin{equation}\label{lanvegin0}
    \mathbf{s}_{l}^{k+1} = \mathbf{s}_{l}^{k} - \frac{1}{2}\mathbf{D}_{l}
    \mathbf{g}(\mathbf{s}_{1:L}^{k})
    + \mathbf{D}_{l}^{\frac{1}{2}}\mathbf{M}_{l}^{\frac{1}{2}} \mathbf{v}_{l}^{k}
\end{equation}

This form is also used in \cite{Grenander94}, \cite{Dostert06}.
The samples are produced by using this first order equation, and
then they are tested in the Metropolis-Hastings scheme.

If we assume the transitions in (\ref{lanvegin0}) as a Wiener
process and take into account the fact that the velocity vector
$\mathbf{v}_{l}$ is independent of the source vector
$\mathbf{s}_{l}$, \cite{Neal93}, then its probability density
function can be set as a multivariate Gaussian as
$p_{\mathbf{v}_{l}}(\mathbf{v}_{l}) =
\left(|\mathbf{M}_{l}|/2\pi\right)^{\frac{1}{2}} \exp \left\{ -
\frac{1}{2} \mathbf{v}_{l}^{T}(t)\mathbf{M}_{l}\mathbf{v}_{l}(t)
\right\}$. We can produce a random sample from this probability
such that $\mathbf{v}_{l} =
\mathbf{M}_{l}^{-\frac{1}{2}}\mathbf{w}_{l}$ where
$\mathbf{w}_{l}$ is a zero-mean Gaussian vector with identity
covariance matrix $\mathcal{N}(\mathbf{w}_{l}|0,\mathbf{I})$. If
we substitute this random sample into (\ref{lanvegin0}), we obtain
the associated Langevin equation
\begin{equation}\label{lanvegin}
    \mathbf{s}_{l}^{k+1} = \mathbf{s}_{l}^{k} - \frac{1}{2}\mathbf{D}_{l}\mathbf{g}(\mathbf{s}_{1:L}^{k})
    + \mathbf{D}_{l}^{\frac{1}{2}} \mathbf{w}_{l}
\end{equation}

Since the random variables for the image pixel intensities are
produced in parallel by using (\ref{lanvegin}), the procedure is
faster than the random walk process adopted in \cite{Kayabol09ip}.
The random walk process produces local random increments
independently from the neighbor pixels and the observations. In
the Langevin sampler, the samples are generated in an interrelated
manner and in terms of the descent of an energy function that
reflects the goodness of the model fit. Once the candidate sample
image is produced by (\ref{lanvegin}), the accept-reject rule is
applied independently to each pixel. In the case of random walk,
we would produce the candidate sample pixel and apply the
accept-reject rule. The sampling of the whole image would be
completed by scanning all the pixels in a sequential order as in
Gibbs sampling. Since each pixel has to wait the update of the
previous pixel, this procedure is very slow. In random walk,
candidate pixels can be produced in parallel but, producing a
candidate sample for the whole image using random walk is not a
reasonable method because hitting the right combination for such a
huge amount of data (i.e. $\approx 10^{5}$) is almost impossible.
By Langevin sampler, the likelihood of approximately hitting the
right combination at any one step is much higher.

After their production, the samples are tested via
Metropolis-Hastings \cite{Hastings} scheme pixel-by-pixel. The
acceptance probability of any proposed sample is defined as
$\min\{\varphi(s_{l,n}^{k+1},s_{l,n}^{k}),1\}$, where
\begin{equation}\label{acptprob}
    \varphi(s_{l,n}^{k+1},s_{l,n}^{k}) \propto
     e^{ - \Delta E(s_{l,n}^{k+1})}\frac{q(s_{l,n}^{k}|s_{l,n}^{k+1})}{q(s_{l,n}^{k+1}|s_{l,n}^{k})}
\end{equation}
where $\Delta E(s_{l,n}^{k+1}) =
E(s_{l,n}^{k+1},s_{(1:L)-l,n}^{k}) - E(s_{1:L,n}^{k})$ and
$E(s_{1:L,n}^{k}) = W(s_{1:L,n}^{k}) + U(s_{l,n}^{k})$. For any
single pixel, $U(s_{l,n})$ can be derived from (\ref{Upl}) and
(\ref{multicauchy}) as
\begin{equation}\label{Using}
    U(s_{l,n}) = \sum_{d=1}^{D} \frac{1+\beta_{l,d}}{2}\log\left[1+\frac{\phi_{d}(s_{l,n},\alpha_{l,d})}{\beta_{l,d}\delta_{l,d}}\right]
\end{equation}

The proposal density $q(s_{l,n}^{k+1}|s_{l,n}^{k})$ is obtained,
from (\ref{lanvegin}), as
\begin{equation}\label{q}
    \mathcal{N} \left(s_{l,n}^{k+1}|s_{l,n}^{k} +
\frac{\tau_{l}^{2}}{2m_{l,n}} g(s_{1:L,n}^{k}),
\frac{\tau_{l}^{2}}{m_{l,n}} \right)
\end{equation}

One cycle of the Metropolis-Hastings algorithm embedded in the
main algorithm, for each source image, is given in Table \ref{MH}.
\begin{table}[!t]
\caption{Metropolis-Hastings algorithm for a source image. $u$:
uniform positive random number in the unit interval; $\mathbf{z}$:
generated sample vector to be tried; $\varphi(z_{n},s_{l,n}^{k})$
: acceptance ratio of the generated sample.}\label{MH}
  \centering
  {{\parbox{7.2cm}{
    \begin{enumerate}
    \item $\mathbf{w}_{l} \sim
        \mathcal{N}(\mathbf{w}_{l}|0,\mathbf{I})$
    \item $\overline{\mathbf{H}}(\mathbf{s}_{l}^{k}) \longleftarrow [\mathrm{diag}
        \left\{\mathbf{H}(\mathbf{s}_{l})
        \right\}_{\mathbf{s}_{l} \longleftarrow \mathbf{s}_{l}^{k}}]^{-1}$
    \item $\mathbf{D}_{l} \longleftarrow 2 [ \overline{\mathbf{H}}(\mathbf{s}_{l}^{k})]^{-1}$
    \item $\mathbf{g}(\mathbf{s}_{1:L}^{k}) \longleftarrow [\nabla_{\mathbf{s}_{l}} E(\mathbf{s}_{1:L})]_{\mathbf{s}_{1:L}
        = \mathbf{s}_{1:L}^{k}}$
    \item produce $\mathbf{z} \longleftarrow \mathbf{s}_{l}^{k} - \frac{1}{2}\mathbf{D}_{l}\mathbf{g}(\mathbf{s}_{1:L}^{k})
    + \mathbf{D}_{l}^{\frac{1}{2}} \mathbf{w}_{l}$ from (\ref{lanvegin}).
    \item for all pixel $n=1,\ldots, N$
            \begin{enumerate}
                \item calculate $\varphi(z_{n},s_{l,n}^{k})$
                \item if $\varphi(z_{n},s_{l,n}^{k})\geq 1$ then $s_{l,n}^{k+1}=z_{n}$\\
                    else produce $u\sim U(0,1)$.
                    \begin{quote}
                    if $u<\varphi(z_{n},s_{l,n}^{k})$ then $s_{l,n}^{k+1}=z_{n}$,\\
                    else $s_{l,n}^{k+1}=s_{l,n}^{k}$
                    \end{quote}
                \item $n+1\longleftarrow$ next pixel.
            \end{enumerate}
    \end{enumerate}
    }}}
\end{table}
\subsubsection{Diffusion Matrix}\label{BolDif}
In this section, we give a method to find an optimum diffusion
matrix $\mathbf{D}$. The method must ensure that the produced
sample $\mathbf{s}_{l}^{k+1}$ comes from the joint conditional
distribution $p(\mathbf{s}_{l}, \mathbf{v}_{l} | \mathbf{y}_{1:K},
\mathbf{s}_{(1:L)-l}, \mathbf{A}, \Theta, \mathbf{M})$ introduced
in (\ref{augpost}). If we write the Taylor expansion of
$E(\mathbf{s}_{l}^{k})$ with the infinitesimal
$\Delta\mathbf{s}_{l}$ and take the expectation of both sides with
respect to the joint density $p(\mathbf{s}_{l}, \mathbf{v}_{l} |
\mathbf{y}_{1:K}, \mathbf{s}_{(1:L)-l}, \mathbf{A}, \Theta,
\mathbf{M})$, we obtain the following equation
\begin{equation}\label{taylorE} \nonumber
    \langle E(\mathbf{s}_{l}+\Delta\mathbf{s}_{l})\rangle  =
    \langle E(\mathbf{s}_{l}) +
    \nabla E(\mathbf{s}_{l})^{T}\Delta\mathbf{s}_{l}   +
    \frac{1}{2}\Delta\mathbf{s}_{l}^{T}\mathbf{H}(\mathbf{s}_{l})\Delta\mathbf{s}_{l}\rangle
\end{equation}
where $\mathbf{H}(\mathbf{s}_{l})$ is the Hessian matrix of
$E(\mathbf{s}_{l})$ with respect to $\mathbf{s}_{l}$. From this
equation, the optimum infinitesimal $\Delta\mathbf{s}_{l}$ is
found as $\Delta\mathbf{s}_{l} = -[\langle
\mathbf{H}(\mathbf{s}_{l})\rangle]^{-1}\langle \nabla
E(\mathbf{s}_{l})\rangle$.

If we also take the expectation of both sides of
(\ref{lanvegin0}), we obtain
\begin{equation}\label{Exlanvegin}
    \langle \mathbf{s}_{l}^{k+1}\rangle = \langle \mathbf{s}_{l}^{k}\rangle - \frac{1}{2}\mathbf{D}_{l} g(\mathbf{s}_{1:L}^{k})
\end{equation}
and comparing $\langle \mathbf{s}_{l}^{k+1}\rangle = \langle
\mathbf{s}_{l}^{k}\rangle + \Delta\mathbf{s}_{l}$ with
(\ref{Exlanvegin}), we write $\mathbf{D}_{l}
g(\mathbf{s}_{1:L}^{k}) = -2[\langle
\mathbf{H}(\mathbf{s}_{l})\rangle]^{-1}\langle \nabla
E(\mathbf{s}_{l})\rangle$. Rather than the expectation of the
inverse of Hessian matrix, we use its diagonal calculated by the
value of $\mathbf{s}_{l}$ at the discrete time $k$ as
\begin{equation}\label{Dder2}
    \mathbf{D}_{l}
g(\mathbf{s}_{1:L}^{k}) = -2[
\overline{\mathbf{H}}(\mathbf{s}_{l}^{k})]^{-1}g(\mathbf{s}_{1:L}^{k})
\end{equation}
where $\overline{\mathbf{H}}(\mathbf{s}_{l}^{k}) = [\mathrm{diag}
\left\{\mathbf{H}(\mathbf{s}_{l})
\right\}_{\mathbf{s}_{l}=\mathbf{s}_{l}^{k}}]^{-1}$ and
$\mathrm{diag}\{.\}$ operator extract the main diagonal of the
Hessian matrix. From (\ref{Dder2}), we can find the diffusion
parameter as \cite{Becker89}:
\begin{equation}\label{Difpar0}
    \mathbf{D}_{l} = 2 [
    \overline{\mathbf{H}}(\mathbf{s}_{l}^{k})]^{-1}.
\end{equation}

This approximation is justified if $\mathbf{H}(\mathbf{s}_{l})$ is
strongly diagonally dominant.
\subsection{Parameters of $t$-distribution } \label{BolReg}
We can write the joint posterior of the parameters $\alpha_{l,d}$,
$\beta_{l,d}$ and $\delta_{l,d}$ such that
$p(\alpha_{l,d},\beta_{l,d},\delta_{l,d}|\mathbf{t}_{l,d},\Theta_{-\{\alpha_{l,d},\beta_{l,d},\delta_{l,d}\}})
= p(\mathbf{t}_{l,d}|\Theta)p(\beta_{l,d})p(\delta_{l,d})$. Using
the likelihood $p(\mathbf{t}_{l,d}|\Theta)$ in (\ref{jointT}) and
the priors of the parameters, we can find the MAP estimates of the
parameters of the $t$-distribution by EM method. Instead of
maximizing the $\log
\left\{p(\mathbf{t}_{l,d}|\Theta)p(\beta_{l,d})p(\delta_{l,d})\right\}$,
we maximize the following function
\begin{eqnarray}\label{emyA}
   \int \log \left\{\frac{p(\mathbf{t}_{l,d}|\Theta)p(\beta_{l,d})p(\delta_{l,d})}
    {p(\nu_{l,d}|\mathbf{t}_{l,d}^{k},\Theta^{k})}\right\}
  p(\nu_{l,d}|\mathbf{t}_{l,d}^{k},\Theta^{k}) d\nu_{l,d}\\ \nonumber
  =  \left \langle \log
  \{p(\mathbf{t}_{l,d}|\Theta)p(\beta_{l,d})p(\delta_{l,d})\}\right\rangle_{\nu_{l,d}|\mathbf{t}_{l,d}^{k},\Theta^{k}} & & \\\nonumber
   - \left \langle \log p(\nu_{l,d}|\mathbf{t}_{l,d}^{k},\Theta^{k})\right\rangle_{\nu_{l,d}|\mathbf{t}_{l,d}^{k},\Theta^{k}} & & \\ \nonumber
\end{eqnarray}
where $p(\nu_{l,d}|\mathbf{t}_{l,d}^{k},\Theta^{k})$ is the
posterior density of the hidden variable $\nu_{l,d}$ conditioned
on parameters estimated in the previous step $k$ and $\left\langle
. \right\rangle_{\nu_{l,d}|\mathbf{t}_{l,d}^{k},\Theta^{k}}$
represents the expectation with respect to
$\nu_{l,d}|\mathbf{t}_{l,d}^{k},\Theta^{k}$. For simplicity,
hereafter we use only $\left\langle . \right\rangle$ to represent
this expectation. The parameter $\nu_{l,d}$ is a hidden (or
latent) variable that changes the scale of the Gaussian density
$\mathcal{N}(\mathbf{t}_{l,d}|0,\delta_{l,d}^{k}\mathbf{I}_{N}/\nu_{l,d})$
and has a Gamma prior
$\mathcal{G}(\nu_{l,d}|\beta_{l,d}^{k}/2,\beta_{l,d}^{k}/2)$. By
exploiting $\nu_{l,d}$, we can define the $t$-distribution as a
scale mixture of Gaussians as in (\ref{jointT}). The second term
on the righthand side of (\ref{emyA}), $ -\langle\log
p(\nu_{l,d}|\mathbf{t}_{l,d}^{k},\Theta^{k})\rangle$, corresponds
to the entropy of the posterior density of $\nu_{l,d}$, and is
independent of the unknowns, and the function
\begin{equation}\label{Q}
    Q(\Theta;\Theta^{k}) =  \left\langle \log \left\{p(\mathbf{t}_{l,d}|\Theta)p(\beta_{l,d})p(\delta_{l,d})\right\}
  \right\rangle.
\end{equation}

The aim is to find the maximum of $Q(\Theta;\Theta^{k})$ with
respect to $\Theta$;
\begin{equation}\label{maxem}
    \Theta^{k+1} = \arg \max_{\Theta} Q(\Theta;\Theta^{k})
\end{equation}

In the E (expectation) step of the EM algorithm, we must calculate
the expectation $\langle .
\rangle_{\nu_{l,d}|\mathbf{t}_{l,d}^{k},\Theta^{k}}$. For this
purpose, we find the posterior density of $\nu_{l,d}$
\begin{equation}\label{postnu}
    \begin{array}{ccc}
      p(\nu_{l,d}|\mathbf{t}_{l,d}^{k},\Theta^{k})  =  p(\mathbf{t}_{l,d}^{k}|\Theta^{k},\nu_{l,d})p(\nu_{l,d}) \\
        =  \mathcal{N}(\mathbf{t}_{l,d}^{k}|0,\delta_{l,d}^{k}\mathbf{I}_{N}/\nu_{l,d}) \mathcal{G}(\nu_{l,d}|\beta_{l,d}^{k}/2,\beta_{l,d}^{k}/2) \\
        =  \mathcal{G}\left(\nu_{l,d}|N/2,\phi_{d}(\mathbf{s}_{l}^{k},\alpha_{l,d}^{k})/\delta_{l,d}^{k}\right) \mathcal{G}\left(\nu_{l,d}|\beta_{l,d}^{k}/2,\beta_{l,d}^{k}/2\right) \\
        =
        \mathcal{G}\left(\nu_{l,d}|\frac{N+\beta_{l,d}^{k}}{2},\frac{\beta_{l,d}^{k}}{2}\left(1+\frac{\phi_{d}(\mathbf{s}_{l}^{k},\alpha_{l,d}^{k})}{\beta_{l,d}^{k}\delta_{l,d}^{k}}\right)\right).
    \end{array}
\end{equation}

The expectation of $\nu_{l,d}$ is
\begin{equation}\label{expectnu}
    \langle\nu_{l,d}\rangle
    = \frac{N+\beta_{l,d}^{k}}{\beta_{l,d}^{k}} \left(1+\frac{\phi_{d}(\mathbf{s}_{l}^{k},\alpha_{l,d}^{k})}{\beta_{l,d}^{k}\delta_{l,d}^{k}}\right)^{-1}
\end{equation}

In the M (maximization) step, (\ref{Q}) is maximized with respect
to $\Theta$. To maximize this function, we alternate among the
variables $\alpha_{l,d}$, $\beta_{l,d}$ and $\delta_{l,d}$. After
taking the logarithms and expectations in (\ref{Q}), the cost
functions for $\alpha_{l,d}$, $\beta_{l,d}$ and $\delta_{l,d}$ are
written as follows
\begin{equation}\label{emal}
    Q(\alpha_{l,d};\Theta^{k}) = - \langle\nu_{l,d}\rangle\frac{\phi_{d}(\mathbf{s}_{l},\alpha_{l,d})}{2\delta_{l,d}}
\end{equation}
\begin{equation}\label{emdel}
    Q(\delta_{l,d};\Theta^{k})  = -\frac{N}{2}\log\delta_{l,d}  -
    \left(
    \langle\nu_{l,d}\rangle\frac{\phi_{d}(\mathbf{s}_{l},\alpha_{l,d})}{2\delta_{l,d}}
    \right)
\end{equation}
\begin{equation}\label{embet}
    \begin{array}{ccc}
      Q(\beta_{l,d};\Theta^{k}) & = & - \log\Gamma(\frac{\beta_{l,d}}{2}) +\left(\frac{N+\beta_{l,d}}{2}-1\right)\langle\log\nu_{l,d}\rangle \\
       &  & + \frac{\beta_{l,d}-1}{2}\log\beta_{l,d} - \frac{N+\beta_{l,d}}{2}\log 2\\
       &  &  - \frac{\langle\nu_{l,d}\rangle\beta_{l,d}^{k}}{2}\left(1+\frac{\phi_{d}(\mathbf{s}_{l},\alpha_{l,d})}{\beta_{l,d}^{k}\delta_{l,d}^{k}}\right) - 0.002\beta_{l,d}
    \end{array}
\end{equation}

The solutions to (\ref{emal}) and (\ref{emdel}) can be easily
found as
\begin{equation}\label{alphaML}
    \alpha_{l,d} = \frac{\mathbf{s}_{l}^{T}\mathbf{G}_{d}^{T}\mathbf{s}_{l}}{\mathbf{s}_{l}^{T}\mathbf{G}_{d}^{T}\mathbf{G}_{d}\mathbf{s}_{l}}
\end{equation}
\begin{equation}\label{deltaML}
    \delta_{l,d} =
    \langle\nu_{l,d}\rangle\frac{\phi_{d}(\mathbf{s}_{l},\alpha_{l,d})}{N}
\end{equation}

The maximization of (\ref{embet}) does not have a simple solution.
It can be solved by setting its first derivative to zero:
\begin{equation}\label{betazero}
\begin{array}{cc}
    - \psi_{1}(\frac{\beta_{l,d}}{2}) + \log\beta_{l,d} + \langle\log\nu_{l,d}\rangle - \langle\nu_{l,d}\rangle \\ + \frac{\beta_{l,d}-1}{\beta_{l,d}} - 0.002= 0
\end{array}
\end{equation}
where $\psi_{1}(.)$ is the first derivative of $\log\Gamma(.)$ and
it is called digamma function.
\subsection{Parameters of the Mixing Matrix} \label{BolMixMat}
We assume that the prior of $\mathbf{A}$ is uniform between 0 and
$\infty$. The conditional density of $a_{k,l}$ is expressed as
$p(a_{k,l}|\mathbf{y}_{1:K}, \Theta_{-a_{k,l}}^{t}) \propto
p(\mathbf{y}_{1:K}|\Theta^{t})$. From (\ref{l}), it can be seen
that the conditional density of $a_{k,l}$ becomes Gaussian. The
parameter $a_{k,l}$ is estimated in each iteration as
\begin{equation}\label{anupda}
    a_{k,l}  =  \frac{1}{\mathbf{s}_{l}^{T}\mathbf{s}_{l}} \mathbf{s}_{l}^{T}(\mathbf{y}_{k}-\sum_{i=1, i\neq
    l}^{L}a_{k,i}\mathbf{s}_{i}) u(a_{k,l})
\end{equation}
where $u(a_{k,l})$ is the unit step function.
\subsection{Adaptive Langevin Sampler Algorithm} \label{BolAlg}
The proposed Adaptive Langevin Sampler algorithm is given in Table
\ref{ALS}. The symbol $\longleftarrow$ denotes analytical update,
the symbol $\longleftarrow_{0}$ denotes update by finding the zero
root and the symbol $\sim$ denotes the update by random sampling.
The sampling of the sources is done by the Metropolis-Hastings
scheme given in Table \ref{MH}. To deal practically with uniformly
distributed positive variables, we assume that they lie in the
range $[0.0001, 1000]$.
\subsubsection{Initialization} \label{ini}
We start the algorithm with the mixing matrix obtained by the
FDCCA (Fourier Domain Correlated Component Analysis)
\cite{Bedini07} method. The initial values of astrophysical maps
are obtained by Least Square (LS) solution with the initial mixing
matrix. The initial values of $\alpha_{l,d}$ can be calculated
directly from image differentials. We initialized the
$\beta_{l,d}^{0}=0$ and found the initial value of $\delta_{l,d}$
by equaling the expectation (\ref{expectnu}) to a constant. In
this study, we take the initial value of this posterior
expectation $1.5$.  So the initial value of $\delta_{l,d}^{0} =
1.5\phi_{d}(\mathbf{s}_{l}^{0},\alpha_{l,d}^{0})/N$
\begin{table}[!t]
\caption{One cycle of Adaptive Langevin Sampler for source
separation. The symbol $\longleftarrow$ denotes analytical update,
the symbol $\longleftarrow_{0}$ denotes update by finding the zero
root and the symbol $\sim$ denotes update by random
sampling.}\label{ALS}
  \centering
  {{\parbox{10.0cm}{
  Find the initial mixing matrix (i.e. FDCCA \cite{Bedini07}).\\
  Find the initial source images using the LS solution.\\
  Initialize the parameters $\alpha_{l,d}^{0}$, $\beta_{l,d}^{0}$ and
  $\delta_{l,d}^{0}$\\
  for all source images, $l=1:L$
  \begin{quote}
    for all directions, $d=1:D$
    \begin{quote}
        $\langle\nu_{l,d}\rangle \longleftarrow
        \frac{N+\beta_{l,d}^{k}}{\beta_{l,d}^{k}} \left(1+\frac{\phi_{d}(\mathbf{s}_{l}^{k},\alpha_{l,d}^{k})}{\beta_{l,d}^{k}\delta_{l,d}^{k}}\right)^{-1}$\\
        $\alpha_{l,d} \longleftarrow
        \frac{\mathbf{s}_{l}^{T}\mathbf{G}_{d}^{T}\mathbf{s}_{l}}{\mathbf{s}_{l}^{T}\mathbf{G}_{d}^{T}\mathbf{G}_{d}\mathbf{s}_{l}}$\\
        $\delta_{l,d} \longleftarrow
            \langle\nu_{l,d}\rangle\frac{\phi_{d}(\mathbf{s}_{l},\alpha_{l,d})}{N}$\\
        $\beta_{l,d} \longleftarrow_{0}  [- \psi_{1}(\frac{\beta_{l,d}}{2}) + \log\beta_{l,d} + \langle\log\nu_{l,d}\rangle - \langle\nu_{l,d}\rangle \\ + \frac{\beta_{l,d}-1}{\beta_{l,d}} - 0.002= 0]$
    \end{quote}

    for all pixels, $n=1:N$
    \begin{quote}
        Using Metropolis-Hastings method in Table
        \ref{MH}\\
        $s_{l,n}^{k+1} \sim \left\{
        p(s_{l,n}|\mathbf{y}_{1:K},\Theta_{-s_{l,n}}^{t})\right\}$
    \end{quote}
  \end{quote}
  for all elements of the mixing matrix, $(k,l)=(1,1):(K,L)$
  \begin{quote}
  $a_{k,l}  \longleftarrow  \frac{1}{\mathbf{s}_{l}^{T}\mathbf{s}_{l}} \mathbf{s}_{l}^{T}(\mathbf{y}_{k}-\sum_{i=1, i\neq
    l}^{L}a_{k,i}\mathbf{s}_{i}) u(a_{k,l})$
  \end{quote}
    }}}
\end{table}
\subsubsection{Stopping Criterion}
We observe the normalized absolute difference between sequential
values of $\mathbf{s}_{l}$ to decide the convergence of the Markov
Chain to an equilibrium. If
$|\mathbf{s}_{l}^{k}-\mathbf{s}_{l}^{k-1}|/|\mathbf{s}_{l}^{k-1}|
\leq 10^{-2}$, we assume the chain has converged to the
equilibrium for $\mathbf{s}_{l}$ and denote this point $T_{l}=k$.
Since we have $L$ parallel chains for $L$ sources, the ending
point of the burn-in period of the whole Monte Carlo chain is
$T_{s} = \max_{l} T_{l}$. We ignore the samples before $T_{s}$. We
keep the iteration going until $T_{e}$ that is the ending point of
the post burn-in period simulation. In the experiments, we have
used 100 iterations after burn-in period, so $T_{e}=T_{s}+100$.
\section{Simulation Results} \label{BolSim}
To test our procedure, we assume nine observation channels with
center frequencies in the range 30--857 GHz, where the dominant
diffuse radiations are the CMB, the galactic synchrotron radiation
and the thermal emission from galactic dust. Except for CMB, these
radiations have unknown emission spectra (that is, the
coefficients $a_{k,l}$ in (\ref{mixeniyi}) are not all known).
Both observation models in (\ref{mixeniyi}) and (\ref{mix3}) are
suitable for our algorithm. In the experiments, we assume the
model in (\ref{mixeniyi}), that is of instantaneous mixtures and
isotropic sky, to be valid. We plan to attack the problems of
space variability and channel-dependent convolutional effects in
the future.

In the sequel, we present astrophysical image separation results
on a comparative basis. The proposed method is denoted as ALS-t
(Adaptive Langevin Sampler-$t$-distribution) and is compared to
four other methods, namely: 1) GS-MRF, which is the MRF model
coupled with Gibbs sampling \cite{Kayabol09ip}; 2) LS, which forms
our initial estimates on the basis of the values of $a_{k,l}$
obtained by FDCCA \cite{Bedini07}; 3) Iterated Conditional Modes
(ICM), which maximizes the conditional pdfs sequentially for each
variable \cite{Rowe99}; 4) ALM-MRF, which is the solution of the
MRF model via Langevin and Metropolis-Hastings schemes
\cite{Kayabol09b}.

\begin{figure} \centering
  \includegraphics[width=3.5in]{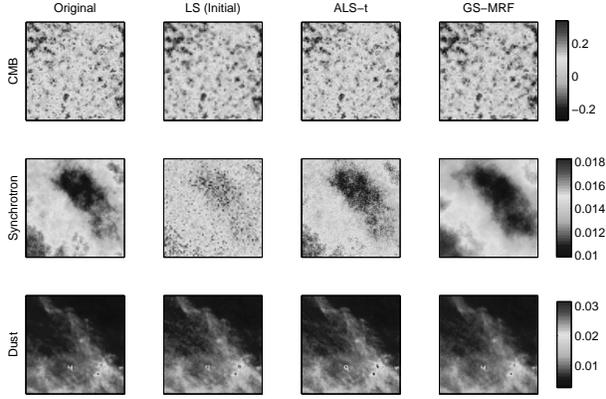}
  \caption{The separated astrophysical images from noisy
observations with the LS, ALS-t and GS-MRF methods. The location
of the patch is $0^{\circ}$ longitude and  $40^{\circ}$ latitude,
out off galactic plane, and has a size $64\times 64$
pixels.}\label{imgeozgur}
\end{figure}

The leftmost column of Fig. \ref{imgeozgur} shows the ground-truth
simulated astrophysical source maps. The remaining columns show
the source maps separated by LS, ALS-t and GS-MRF, respectively.
The sky patch used for this experiment is centered at $0^{\circ}$
longitude and $40^{\circ}$ latitude in galactic coordinates and
has a size of $7.3\times 7.3$ square degrees in the celestial
sphere, discretized in a $64\times 64$ pixel map.

\begin{figure}[]
 \centerline{\includegraphics[width=3.0in]{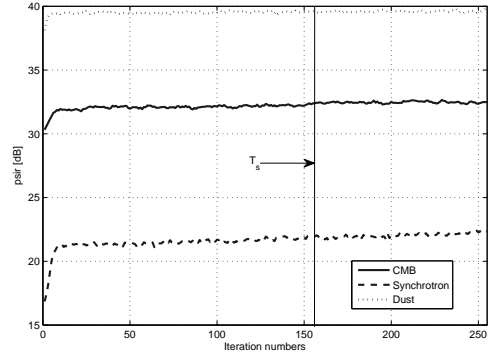}}
\caption{The PSIRs of the sources in the pixel domain as a
function of iteration number. The vertical line signifies the
starting point $T_{s}$.}
 \label{psirit}
\end{figure}

The Peak Signal-to-Interference Ratio (PSIR) is used as a
numerical performance indicator. The PSIR can be calculated if the
ground-truth is known, which is the case in our work since all sky
components are simulated. For this patch, the algorithm converges
after 155 iterations and uses a total of  255 iterations to reach
the solution (see Fig. \ref{psirit}). We compare the results with
the ones of LS, ICM, GS-MRF and ALM-MRF.

Table \ref{Planckno} lists the PSIR values and the process times.
The simulations are run on a Core2 CPU 1.86 GHz PC. The process
time of ALS-t is much shorter than that of the GS-MRF. The
execution time of ALS-t is two orders of magnitude smaller than
that of GS-MRF. The PSIR values of ALS-t are also over those of
LS, ALM-MRF, GS-MRF and ICM, especially for synchrotron, and
furthermore, the smoothing degradation of ICM on the synchrotron
component is not observed in the proposed method , Fig.
\ref{imgeozgur}.
\begin{table}
  \caption{The PSIR (dB) values of the separated components and the process time of the algorithms in minutes.}
  \label{Planckno}
  \centering
\begin{tabular}{|c||c|c|c|c|}
   \hline
   & CMB \hspace{8pt}& Synchrotron & Dust\hspace{7pt} & time \\
  \hline
   LS                           & 30.69 & 15.03 & 37.37 & 1.32e-4\\
   ICM                          & 26.27 & 17.64 & 35.30 & 0.31\\
   GS-MRF, \cite{Kayabol09ip}   & 27.81 & 22.33 & 38.93 & 226.72\\
   ALM-MRF, \cite{Kayabol09b}   & 27.91 & 20.88 & 36.41 & 2.86\\
   ALS-t                        & 33.45 & 26.21 & 40.51 & 1.65\\
  \hline
\end{tabular}
\end{table}
\begin{table}
  \caption{The PSIR improvements (dB) with respect to initial LS solution.}
  \label{Planckno2000}
  \centering
\begin{tabular}{|c||c|c|c|}
   \hline
   & CMB \hspace{8pt}& Synchrotron & Dust\hspace{7pt} \\
  \hline
   $(0^{\circ},40^{\circ})$     & 3.01 & 10.01 & 4.08 \\
   $(20^{\circ},0^{\circ})$     & 1.80 &  4.54 & 1.78 \\
  \hline
\end{tabular}
\end{table}
\begin{figure}[ht] \centering
  \includegraphics[width=3.6in]{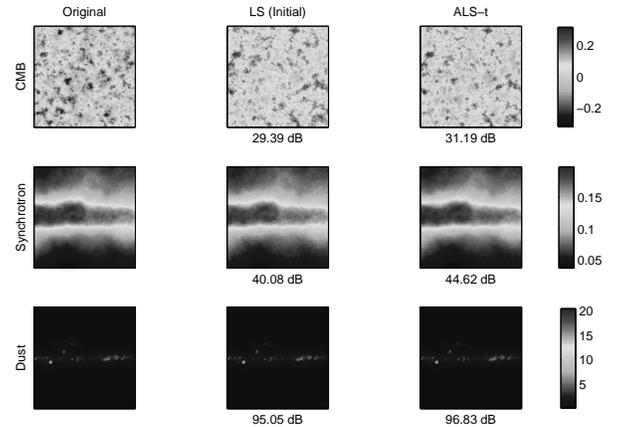}
  \caption{The separated astrophysical images from noisy
observations with the LS and ALS-t methods. The location of the
patch is $20^{\circ}$ longitude and  $0^{\circ}$ latitude,
galactic plane, and has a size $128\times 128$
pixels.}\label{f2000}
\end{figure}

We have also run the algorithm for $128\times 128$ pixels patches
centered at $(0^{\circ},40^{\circ})$ and $(20^{\circ},0^{\circ})$.
Fig. \ref{f2000} shows the results for the patch
$(20^{\circ},0^{\circ})$. In that patch, the relative intensity of
CMB is the weakest one. The PSIR values of the estimates of LS and
ALS-t are written under the maps and the PSIR improvements are
listed for that patch and for the patch $(0^{\circ},40^{\circ})$
in Table \ref{Planckno2000}. The total time of the ALS-t algorithm
for the $128\times 128$ size patches is about $5.31$ minutes.

We also use an alternative performance criterion, defined in the
spherical harmonic (frequency), $\ell$, domain, since the angular
power spectrum is relevant to astrophysics. If we decompose a CMB
map on spherical harmonics, the complex coefficients, $c_{\ell m}$
($\ell = 0, 1,2,\ldots$, $m \in [-\ell,\ell]$), define the angular
power spectrum, $C(\ell)$, as the average $C(\ell) =
\frac{1}{2\ell+1} \sum_{m=-\ell}^{\ell} c_{\ell m}c_{\ell
m}^{\ast}$.

\begin{figure}%
\centering \subfigure[CMB, patch
$(0^{\circ},40^{\circ})$]{\includegraphics[width=1.6in]{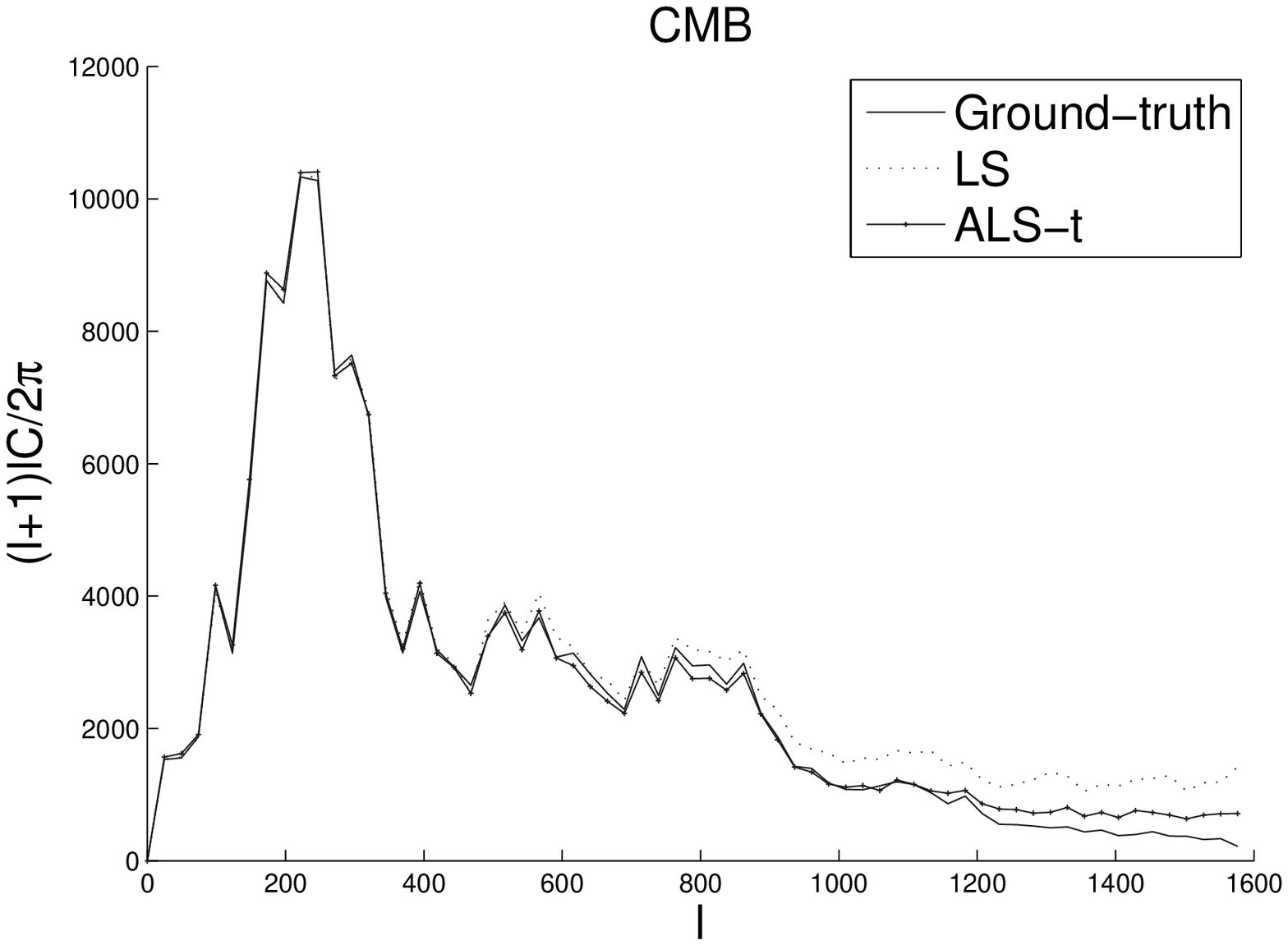}
\label{fig_first_case}} \hfil \subfigure[CMB, patch
$(20^{\circ},0^{\circ})$]{\includegraphics[width=1.6in]{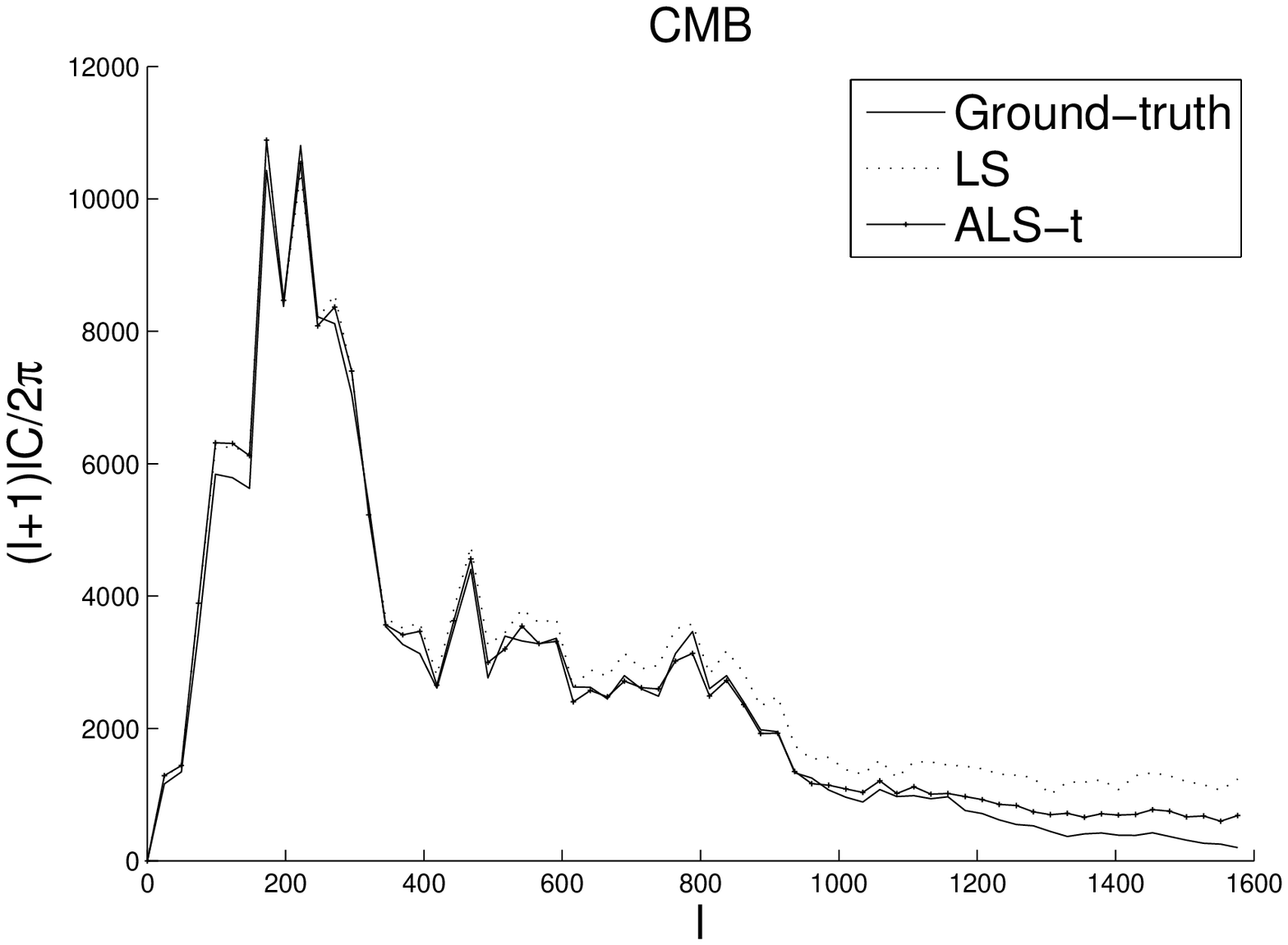}
\label{fig_first_case}}\\

\subfigure[Synchrotron, patch
$(0^{\circ},40^{\circ})$]{\includegraphics[width=1.6in]{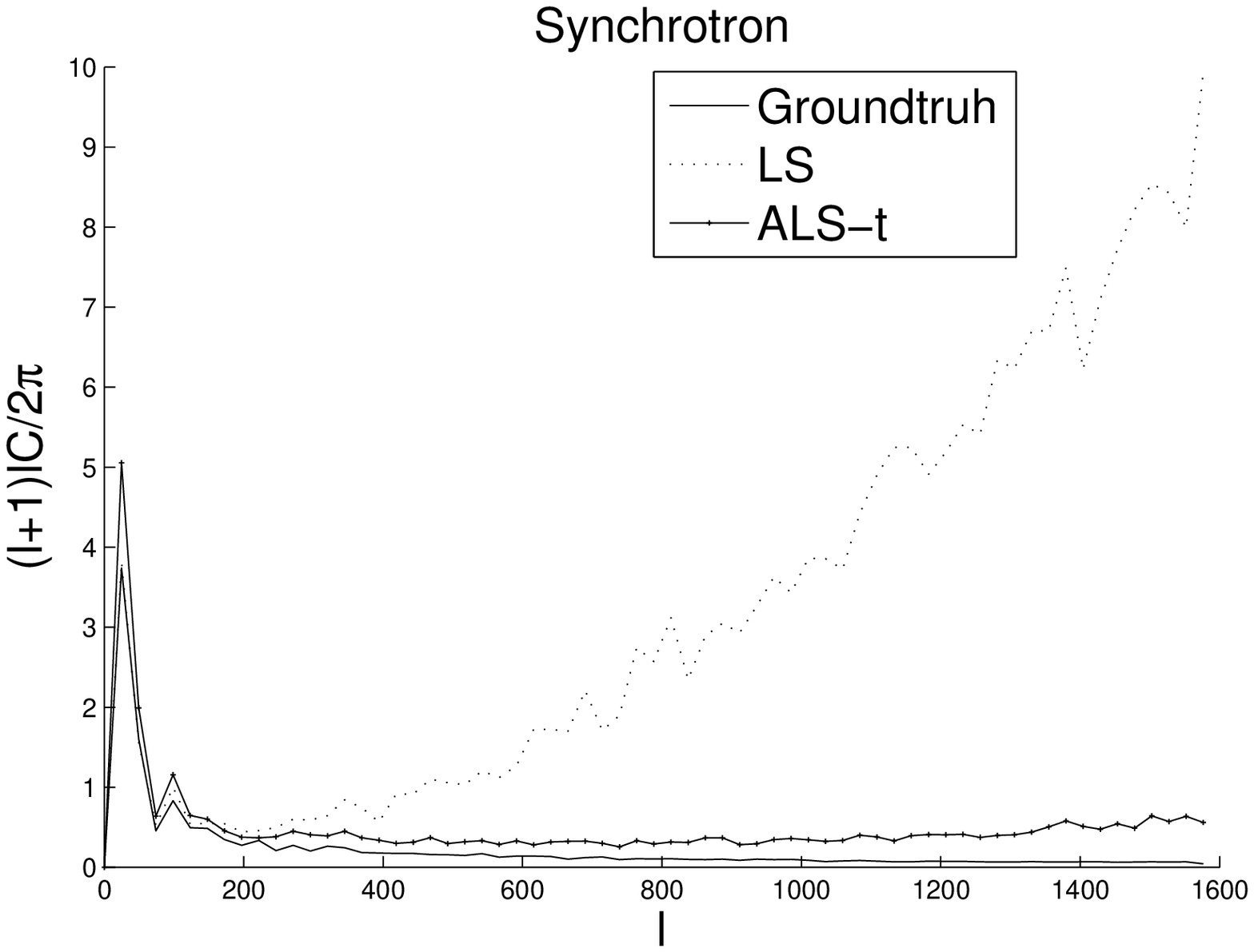}
\label{fig_second_case}} \hfil \subfigure[Synchrotron, patch
$(20^{\circ},0^{\circ})$]{\includegraphics[width=1.6in]{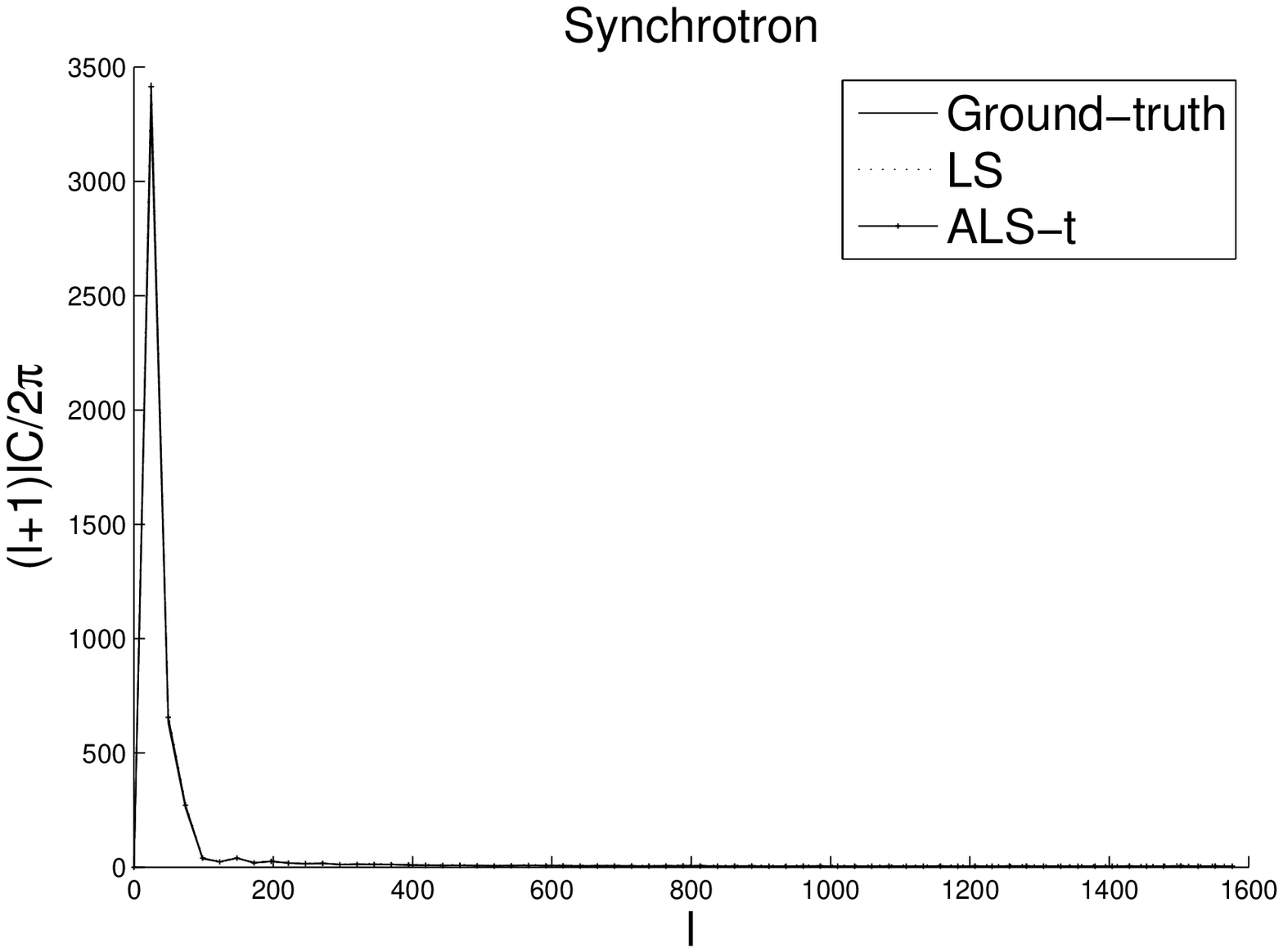}
\label{fig_second_case}}\\
\subfigure[Dust, patch
$(0^{\circ},40^{\circ})$]{\includegraphics[width=1.6in]{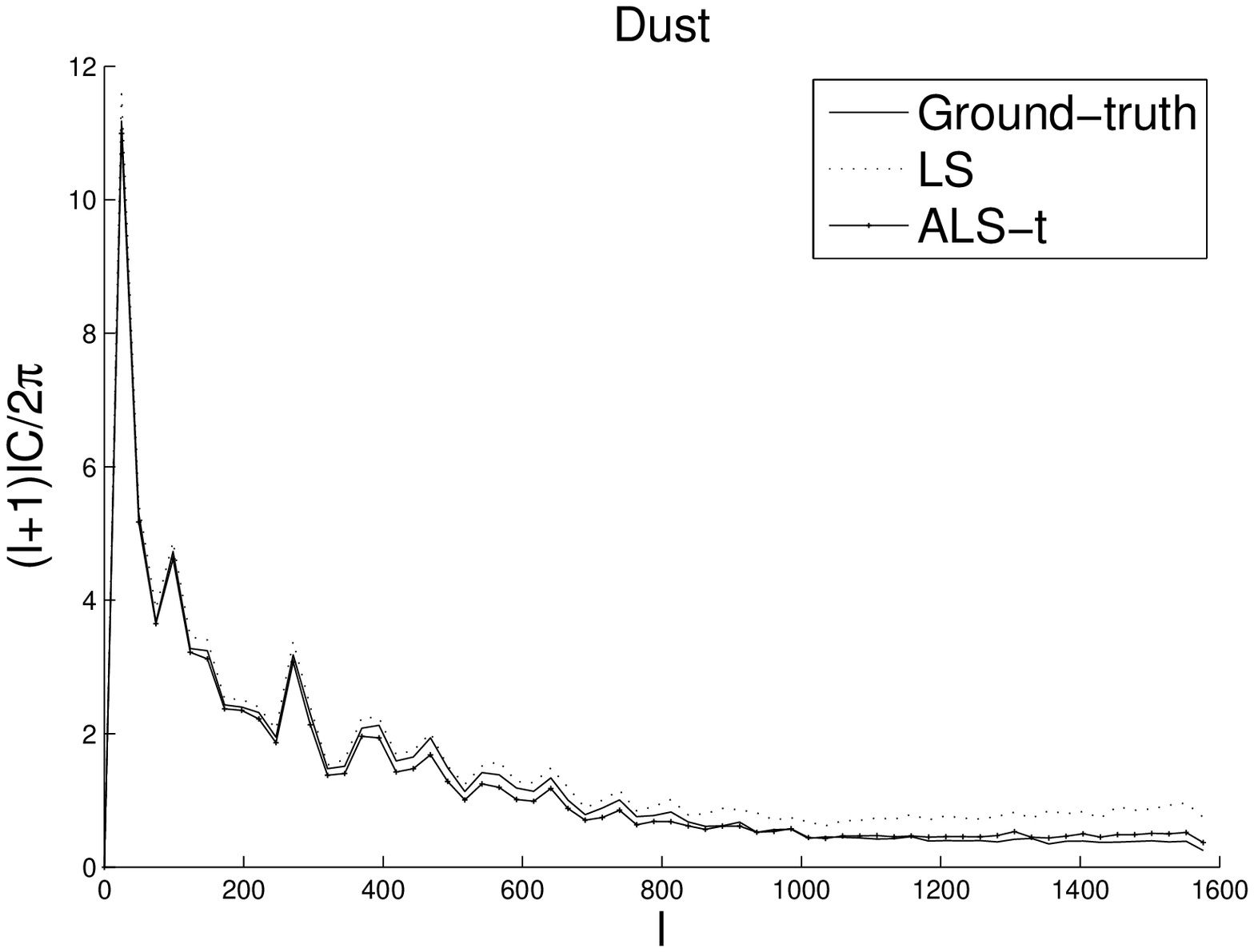}
\label{fig_second_case}} \hfill \subfigure[Dust, patch
$(20^{\circ},0^{\circ})$]{\includegraphics[width=1.6in]{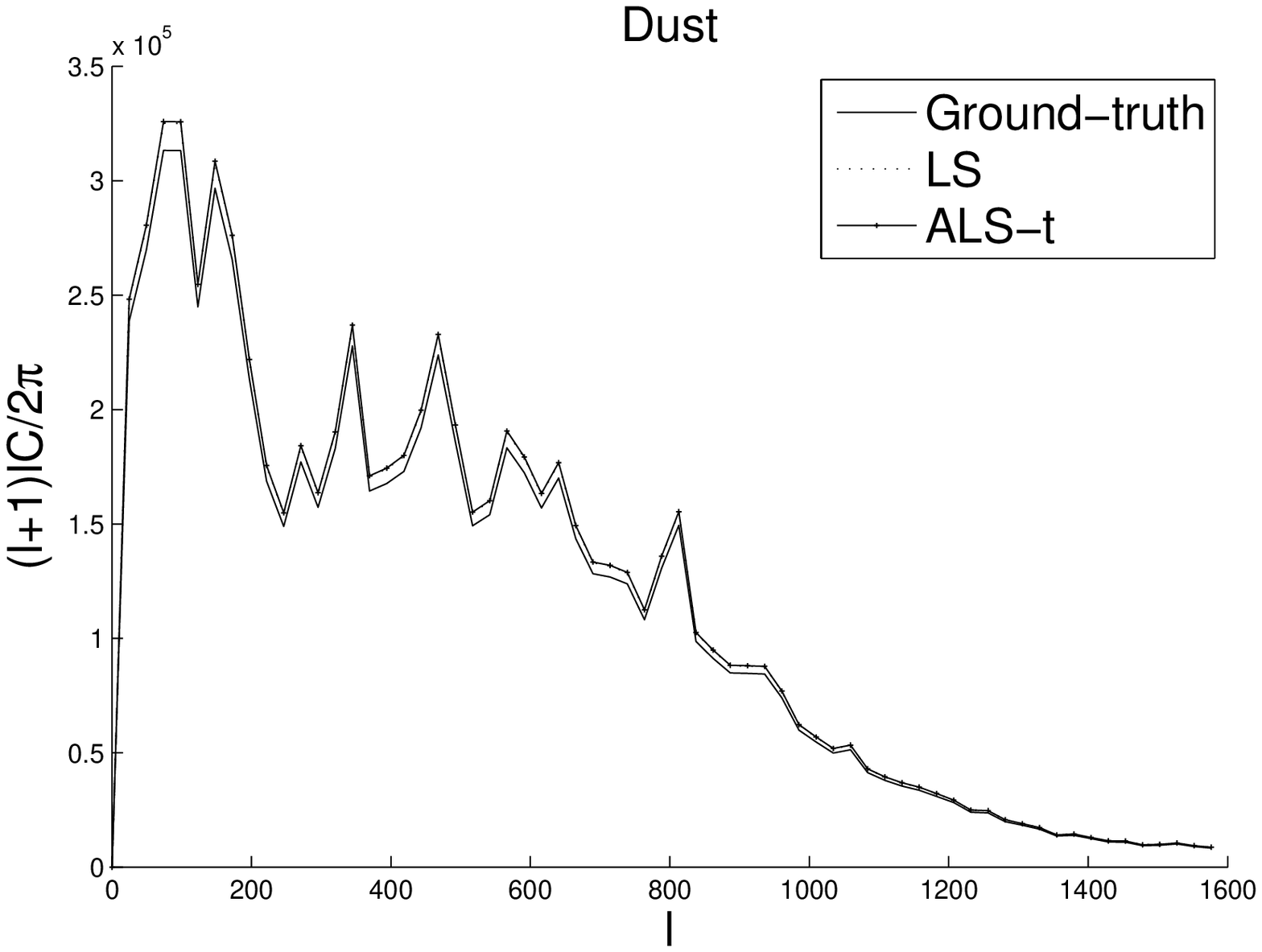}
\label{fig_second_case}}%
 \caption{Ground-truth and
estimated angular power spectrums for patches
$(0^{\circ},40^{\circ})$ and $(20,0)$. The ground-truth spectrum
(solid line), spectrum of LS solution (dot line) and spectrum of
ALS-t solution (solid line marked with +).} \label{p0040}
\end{figure}

In Fig. \ref{p0040}, we plot the standard power spectrum,
$\overline{C}(\ell)$, defined as $\overline{C}(\ell) =
(\ell+1)\ell C(\ell)/2\pi$ of the original and the reconstructed
sources in the two patches considered. In order to compare
different methods, we also introduce the Peak
Signal-to-Interference Ratio in the $\ell$-domain defined as
\begin{equation}\label{snrl}
    PSIR_{spec}=20\log\left(\frac{\sqrt{\sqrt{N}/2+1}\times \max (\overline{C}(\ell))}{||\overline{C}(\ell)-\widehat{\overline{C}}(\ell)||}\right)
\end{equation}
where $\widehat{\overline{C}}(\ell)$ is the estimated power
spectrum.

In the off-galactic patch considered in Fig. \ref{p0040}, the
intensity of synchrotron is very low and the LS solution for
synchrotron is contaminated too much by noise. The estimated CMB
and the dust spectrums by ALS-t follow the ground-truth spectrum
better than the LS one, especially in the high frequency regions.
For the patch $(20^{\circ},0^{\circ})$, synchrotron and dust are
estimated adequately by LS, but the LS estimate of CMB is improved
by ALS-t. The related PSIR$_{spec}$ values are presented in Table
\ref{Tsir}.

\begin{table}
  \caption{The PSIR$_{spec}$ (dB) values of the separated components, in the annular frequency domain.}
  \label{Tsir}
  \centering
\begin{tabular}{|c||c|c|c||c|c|c|}
  \multicolumn{1}{}{} &   \multicolumn{3}{c}{$0^{\circ}$,$40^{\circ}$} &   \multicolumn{3}{c}{$20^{\circ}$,$0^{\circ}$}    \\
   \hline
   & CMB \hspace{3pt}& Synch. & Dust\hspace{2pt} & CMB \hspace{3pt}& Synch. & Dust\hspace{2pt} \\
  \hline
   LS                           & 30.33 &  2.65 & 37.76 & 31.59 & 45.69 & 39.35 \\
   ALS-t                        & 35.87 & 26.23 & 40.69 & 34.53 & 46.98 & 38.93 \\
  \hline
\end{tabular}
\end{table}

We have observed that the estimated regression parameter
$\alpha_{l,d}$ is quite isotropic for all the maps. For the CMB
map in the $(0^{\circ},40^{\circ})$ patch, its value is about
$0.88$ for all $d$. In the same patch, the values of the parameter
$\alpha_{l,d}$ for synchrotron and dust are $0.99$. These results
show us that the CMB radiation is spatially less correlated than
the other radiation sources. We assume the parameter $\beta_{l,d}$
is isotropic and estimate a single value for each direction. The
EM estimation of $\beta_{l,d}$ depends too much on its prior and
initial value. We have allowed the parameter $\delta_{l,d}$ to be
anisotropic, but at the end of the estimation steps we have found
that it is almost isotropic for all radiation maps.

\section{Conclusion and Future Work} \label{BolVargi}

We have developed a Bayesian source separation algorithm for
astrophysical images where the MCMC samples are generated through
the Langevin stochastic equation. The proposed algorithm provides
two orders of magnitude computational economy vis-\`{a}-vis the
Gibbs sampling approach. In addition, it generates better source
separation as compared to all its competitors, i.e., LS, ICM,
ALM-MRF and GS-MRF methods measured in terms of PSIR in the pixel
domain and PSIR$_{spec}$ in the annular frequency domain. The
algorithm can reconstruct the high frequency regions of the power
spectrums with higher fidelity. A byproduct of this approach is
the capability to estimate the parameters of the $t$-distribution
image priors. Although the proposed ALS-t method takes longer than
either ICM or LS methods, its superior performance by far
outweighs this disadvantage, and furthermore the algorithm lends
itself to parallel processing. To improve the algorithm
performance, non-stationary image priors, more efficient
discretization time step and diffusion matrix can be investigated
in the future. Another point to obtain a beneficial algorithm
might be the usage of more than one MH-steps, because the EM
algorithm which estimates parameters converges faster than the
Monte Carlo sampling scheme.

Our new goal is the application of the proposed algorithm to
whole-sky maps. To avoid the difficulties inherent in this
problem, we plan to use the "nested numbering" structure provided
by the HEALPix \cite {Gorski05} package. In this format, we can
reach the indexes of the eight neighbors of each pixel on the
sphere. To calculate the pixel differences, we will implement a
gradient calculation method on the sphere by taking the
non-homogeneous spatial distances between the pixels on the sphere
into consideration.

Other issues to be addressed are the channel-dependent blurring
effects of the antennas and the non-stationary nature of noise. We
have to reformulate the source separation problem without these
simplifying assumptions on the observations. Finally, a
pixel-based estimation error is being analyzed with the goal of
defining a stopping criterion for the algorithm.


%
%
%
%

\section*{Acknowledgment}
The simulated source maps (see \cite{Bedini05}) were provided by
the Planck technical working group on diffuse component separation
(WG2.1). Some of the results in this paper have been derived using
the HEALPix package \cite {Gorski05}.

\begin{biography}{Koray Kayabol}
(S'03, M'09) was born in Sakarya, Turkey in 1977. He received the
B.Sc., M.Sc. and Ph.D. degrees in electrical\&electronics
engineering from Istanbul University, Istanbul, Turkey in 1997,
2002 and 2008, respectively.

He was a research assistant in Electrical \& Electronics Eng.
Dept. between 2001 and 2008. Since 2008, he has been with the
ISTI-CNR, Pisa, Italy as a postdoctoral researcher. His research
interests include Bayesian image processing and statistical image
models.
\end{biography}


\begin{biography}{Ercan Kuruo\u{g}lu}
(SM'06, M'98) was born in 1969 in Ankara, Turkey. He received his
PhD degree from the University of Cambridge in 1998.

He joined the Xerox Research Center, Europe in Cambridge in 1998.
He was an ERCIM Fellow in 2000 in INRIA-Sophia Antipolis, France.
In January 2002, he joined ISTI-CNR, Pisa. He was a visiting
professor in Georgia Tech-Shanghai in Autumn 2007. He is currently
a Senior Researcher at ISTI-CNR. His research interests are in the
areas of statistical signal and image processing and information
and coding theory with applications in astrophysics, geophysics,
bioinformatics and telecommunications.

He was an Associate Editor for IEEE Transactions  on Signal
Processing in 2002-2006 and for IEEE Transactions on Image
Processing in 2005-2009 and is  in the editorial boards of Digital
Signal Processing: a Review Journal and EURASIP Journal on
Advances in Signal Processing. He acted as the technical chair for
EUSIPCO 2006. He is a member of the IEEE Technical Committee on
Signal Processing Theory and Methods.
\end{biography}

\begin{biography}{Jos\'e Luis Sanz}
received the Ph.D. degree in theoretical physics from Universidad
Autonoma de Madrid, Spain, in 1976. He was a M.E.C. Postdoctoral
Fellow at the Queen Mary College, London, U.K., during 1978. He is
currently at the Instituto de F\'\i sica de Cantabria, Santander,
Spain, as UC Professor on Astronomy since 1987.

His research interests are in the areas of Cosmic Microwave
Background astronomy (anisotropies, non-Gaussianity),
extragalactic point sources and clusters of galaxies
(blind/non-blind detection, estimation, statistics) as well as the
development of techniques in signal processing (wavelet design,
linear/non-linear filters, time-frequency, sparse representations)
and application of such tools to astronomical data.
\end{biography}

\begin{biography}{B\"{u}lent Sankur}
has received his B.S. degree in Electrical Engineering at Robert
College, Istanbul, and completed his graduate studies at
Rensselaer Polytechnic Institute, New York, USA. His research
interests are in the areas of Digital Signal Processing, Image and
Video Compression, Biometry, Cognition and Multimedia Systems. He
has established a Signal and Image Processing laboratory and has
been publishing 150 journal and conference articles in these
areas.

Since then he has been at Bo\u{g}azi\c{c}i (Bosporus) University
in the Department of Electric and Electronic Engineering. He has
held visiting positions at University of Ottawa, Technical
University of Delft, and Ecole Nationale Supérieure des
Télécommications, Paris. He also served as a consultant in several
private and government institutions.

Prof. Sankur is serving in the editorial boards of three journals
on signal processing. He was the chairman of ICT'96: International
Conference on Telecommunications and EUSIPCO'05: The European
Conference on Signal Processing as well as technical chairman of
ICASSP'00.
\end{biography}

\begin{biography}{Emanuele Salerno}
graduated in electronic engineering from the University of Pisa,
Italy, in 1985. In 1987, he joined the Italian National Research
Council as a full-time researcher. At present, he is a senior
researcher at the Intitute of Information Science and Technologies
in Pisa, Signals and Images Laboratory.

He has been working in applied inverse problems, image
reconstruction and restoration, microwave nondestructive
evaluation, and blind signal separation, and held various
responsibilities in research programs in nondestructive testing,
robotics, numerical models for image reconstruction and computer
vision, and neural network techniques in astrophysical imagery. He
has been supervising various theses in Computer Science,
Electronic and Communications Engineering, and Physics, and, since
1996, has been teaching courses at the University of Pisa.

Dr. Salerno is a member of the Italian society for information and
communication technology (AICT-AEIT).
\end{biography}

\begin{biography}{Diego Herranz}
received the B.S. degree in 1995 and the M.S. degree in physics
from the Universidad Complutense de Madrid, Madrid, Spain, in 1995
and the Ph.D. degree in astrophysics from Universidad de
Cantabria, Santander, Spain, in 2002. He was a CMBNET Postdoctoral
Fellow at the Istituto di Scienza e Tecnologie dell'Informazione
``A. Faedo'' (CNR), Pisa, Italy, from 2002 to 2004. He is
currently at the Instituto de F\'\i sica de Cantabria, Santander,
Spain, as UC Teaching Assistant.

His research interests are in the areas of Cosmic Microwave
Background astronomy and extragalactic point source statistics as
well as the application of statistical signal processing to
astronomical data, including blind source separation, linear and
nonlinear data filtering, and statistical modeling of heavy-tailed
processes.
\end{biography}






\end{document}